\documentclass{emulateapj}

\newcommand{\rev}[1]{}
\newcommand{\corrected}[1]{#1}

\newcommand{\ph}{\phantom{0}}

\newcommand{\AAp}{\aap}
\newcommand{\ApJ}{\apj}
\newcommand{\ApJS}{\apjs}
\newcommand{\AJ}{\aj}
\newcommand{\ARevEPS}{Ann. Rev. Earth Planet. Sci.}
\newcommand{\JGR}{\jgr}

\newcommand{\be}{\begin{equation}}
\newcommand{\ee}{\end{equation}}
\newcommand{\bd}{\begin{displaymath}}
\newcommand{\ed}{\end{displaymath}}
\newcommand{\bea}{\begin{eqnarray}}
\newcommand{\eea}{\end{eqnarray}}
\newcommand{\mum}{\,\mu\hbox{m}}
\newcommand{\m}{\,\mbox{m}}
\newcommand{\mm}{\,\hbox{mm}}
\newcommand{\cm}{\,\mbox{cm}}

\newcommand{\km}{\,\mbox{km}}
\newcommand{\AU}{\,\mbox{AU}}
\newcommand{\g}{\,\mbox{g}}

\newcommand{\K}{\,\mbox{K}}
\newcommand{\D}{\,\mbox{d}}

\shortauthors{Krivov et al.}
\shorttitle{Debris disk models and planetesimal properties}

\begin{document}


\title{Collisional and Thermal Emission Models of Debris Disks:\\
       Towards Planetesimal Population Properties
      }
\author{Alexander V. Krivov,
        Sebastian M{\"u}ller,
        Torsten L{\"o}hne, and 
        Harald Mutschke}
\affil{Astrophysikalisches Institut und Universit{\"a}tssternwarte,
       Friedrich Schiller University Jena,
       Schillerg{\"a}{\ss}chen~ 2--3, 07745 Jena, Germany;
       krivov@astro.uni-jena.de}

\begin{abstract}
Debris disks around main-sequence stars are believed to derive
from planetesimal populations that have accreted at
early epochs and survived possible planet formation processes.
While debris disks must contain solids in a broad range of sizes~---
from big planetesimals down to tiny dust grains~--- debris disk observations
are only sensitive to the dust end of the size distribution.
Collisional models of debris disks are needed
to ``climb up'' the ladder of the collisional cascade, from dust
towards parent bodies, representing the main mass reservoir of the disks.
We have used our collisional code to
generate five disks around a sun-like star, assuming
planetesimal belts at 3, 10, 30, 100, and $200\AU$ with 10 times the Edgeworth-Kuiper-belt mass
density, and to evolve them for 10~Gyr.
Along with an appropriate scaling rule, this effectively yields a
three-parametric set of reference disks (initial mass, location of
planetesimal belt, age).
For all the disks, we have generated
spectral energy distributions (SEDs),
assuming homogeneous spherical astrosilicate dust grains.
A comparison between generated and actually observed SEDs
yields estimates of planetesimal properties (location, total mass etc.).
As a test and a first application of this approach,
we have selected five disks around sun-like stars with well-known SEDs.
In four cases, we have reproduced the data
with a linear combination of two disks from the grid
(an ``asteroid belt'' at $3\AU$ and an outer ``Kuiper belt'');
in one case a single, outer component was sufficient.
The outer components are compatible with ``large Kuiper belts''
of 0.2--50 earth masses (in the bodies up to $100\km$ in size)
with radii of $100$--$200\AU$.
\end{abstract}


\section{Introduction}

Since the IRAS discovery of the excess infrared emission around Vega by 
\citet{aumann-et-al-1984}, infrared surveys with IRAS, ISO, Spitzer, and
other space-based and ground-based telescopes 
have shown the Vega phenomenon to be common for main-sequence stars
\cite[e.g.][]{meyer-et-al-2004,beichman-et-al-2005,najita-williams-2005,rieke-et-al-2005,
bryden-et-al-2006,siegler-et-al-2006, su-et-al-2006, 
trilling-et-al-2007,hillenbrand-et-al-2008,trilling-et-al-2008}.
The observed excesses are attributed to 
circumstellar disks of second-generation dust,
sustained by numerous planetesimals in orbit around the stars.
Jostling collisions between planetesimals grind them all the way down to
smallest dust grains which are then blown away by stellar radiation.
While the bulk of such a
debris disk's mass is hidden in invisible parent bodies,
the observed luminosity is dominated by small particles at dust sizes.
Hence the studies of dust emission have a potential to shed light onto
the properties of parent planetesimal populations as well
as planets that may shape them and,
ultimately, onto the evolutionary history of circumstellar planetary systems.

However, there is no direct way to infer
the properties of invisible planetesimal populations
from the observed dust emission. Dust and planetesimals can only be linked
through models.
First, dynamical models can be used
to predict, for a given planetesimal family (mass, location, age, etc.), the distribution of dust.
Such models have become available in recent years
\citep[e.g.][]{thebault-et-al-2003,krivov-et-al-2006,thebault-augereau-2007,wyatt-et-al-2007,loehne-et-al-2007}.
After that, standard thermal emission models will describe the resulting dust emission.
Comparison of that emission to the one actually observed would then reveal the probable properties of
underlying, dust-producing planetesimal families.

In this paper, we follow this approach
and generate a set of hypothetical debris disks around G2 dwarfs
with different ages (10~Myr -- 10~Gyr), assuming debris dust to stem from planetesimal
belts with different initial masses at different distances from the
central star.
For every set of these parameters, we simulate steady-state dust distributions 
 with our collisional code
\citep{krivov-et-al-2005,krivov-et-al-2006,loehne-et-al-2007}.
This is different from a traditional, ``empirical'' approach,
in which dust distributions are postulated, usually in form of power laws, 
parameterized by ranges and exponents that play a role of fitting parameters
\citep[e.g.][]{wolf-hillenbrand-2003}.
Interestingly, replacing formal dust distributions with those coming out
of dynamical modeling does not increase the number of fitting parameters.
Just the opposite: the number of parameters reduces and those parameters that
we keep free all have clear astrophysical meaning.
The most important are
location of a parent planetesimal belt and its current mass \citep{wyatt-et-al-2007}.

Having produced a set of model debris disks,  we compute thermal emission fluxes in a wide 
range of wavelengths from mid-infrared to millimeter. In so doing, we completely abandon simple blackbody
or modified blackbody calculations and solve a thermal balance equation
instead. At this stage, we assume compact spherical grains composed of astronomical silicate
\corrected{\citep{laor-draine-1993}}
and employ standard Mie calculations to compute dust opacities.
Although this is still a noticeable simplification, it represents a natural
step towards considering realistic materials and using more involved methods of light
scattering theory that we leave for subsequent papers.

As a test and a first application of the results,
we re-interprete available observational data on a selection of disks
around sun-like stars with well known spectral energy distributions (SEDs).

This paper is organized as follows.
Section 2 describes the dynamical and thermal emission models.
In section~3, a set of reference disks is introduced and the model parameters
are specified.
Section~4 presents the modeling results for this set of disks: size and spatial
distribution of dust, dust temperatures, and the generated SEDs.
Application to selected observed disks is made in section 5.
Section~6 summarizes the paper.

\section{Model}
\subsection{Dynamical model}

To simulate the dust production by the planetesimal belt and
the dynamical evolution of a disk, we use our collisional
code (ACE, Analysis of Collisional Evolution).
The code numerically solves the Boltzmann-Smoluchowski kinetic equation
to evolve a disk of solids in a broad range of sizes
(from smallest dust grains to planetesimals),
orbiting a primary in nearly Keplerian orbits
(gravity + direct radiation pressure + drag forces) and
experiencing disruptive and erosive (cratering) collisions.
Collision outcomes are simulated with available material- and size-dependent scaling
laws for fragmentation and dispersal in both strength and gravity regime.
The current version implements a 3-dimensional kinetic model, with masses,
semimajor axes, and eccentricities as phase space variables.
This approach automatically enables a study of the simultaneous evolution of
mass, spatial, and velocity distribution of particles.
The code is fast enough to easily follow the evolution of a debris disk over
Gyr timescales.
A detailed description of our approach, its numerical implementation, and astrophysical
applications can be found in our previous papers
\citep{krivov-et-al-2000b,krivov-et-al-2005,krivov-et-al-2006,loehne-et-al-2007}.

\subsection{Thermal emission model}

For spherical dust grains with radius $s$ and temperature $T_{\mathrm{g}}$
we can calculate their
distance $r$ to the star under the assumption of thermal equilibrium as
\bea
  r = \frac{R_*}{2} \sqrt{ \frac{\int_0^{\infty} \D\lambda \; Q_{\lambda}^{\mathrm{abs}}(s) 
                                     F_{\lambda, *}(T_*)}
                                {\int_0^{\infty} \D\lambda \; Q_{\lambda}^{\mathrm{abs}}(s) 
                                     B_{\lambda}(T_{\mathrm{g}})} }
 \label{equ:r} .
\eea
Here, $R_*$ denotes the radius and $F_{\lambda, *}(T_*)$ the flux
of the star with an effective temperature $T_*$
and $B_{\lambda}(T_{\mathrm{g}})$ the Planck function.
The absorption efficiency $Q_{\lambda}^{\mathrm{abs}}(s)$
is a function of wavelength $\lambda$ and particle size.

We now consider a rotationally symmetric dust disk at a distance $D$ from the observer.
Denote by $ N(r,s)$
the surface number density of grains with radius $s$ at a distance $r$ from the star,
so that $N(r,s)ds$ is the number of grains with radii $[s,s+ds]$ in a
narrow annulus of radius $r$, divided by the surface area of that annulus.
Then the specific flux emitted from the entire disk at a given wavelength can be calculated as
\bea
  F_{\lambda, \mathrm{disk}}^{\mathrm{tot}}
            & = & \int \D r \int \D s  \; F_{\lambda, \mathrm{disk}}(r,s) \label{equ:F1}\\
            & = & \frac{2\pi^2}{D^2} \int \D T_g \, r(T_g) \, \frac{\D r(T_g)}{\D T_g} \int \D s \; s^2 
\;\times\nonumber\\
            &   & \times \; N(r,s) \,Q_{\lambda}^{abs}(s) \, B_{\lambda}(T_g)
  \label{equ:F} .
\eea

\section{Reference disks}

\subsection{Central star}

The parameters of the central star (mass and photospheric spectrum)
affect both the dynamics of solids
(by setting the scale of orbital velocities and
determining the radiation pressure strength)
and their thermal emission
(by setting the dust grain temperatures).
We take the Sun (a G2V dwarf with a solar metallicity)
as a central star and calculate its photospheric spectrum with the NextGen
grid of models \citep{hauschildt-et-al-1999}.

\subsection{Forces}

In the dynamical model, we include central star's gravity and
direct radiation pressure. We switch off the drag forces (both the 
Poynting-Robertson and stellar wind drag), which are of little
importance for the optical depths in the range from $\sim 10^{-5}$ to $\sim 10^{-3}$)
considered here \citep{artymowicz-1997,krivov-et-al-2000b,wyatt-2005}.

\subsection{Collisions}

The radii of solids in every modeled disk cover the interval from $0.1\mum$ to $100\km$.
The upper limit of $100\km$ is justified by the fact that planetesimal accretion
models predict larger objects to have a steeper size distribution and thus to contribute
less to the mass budget of a debris disk \citep[e.g.][]{kenyon-luu-1999b}.
To describe the collisional outcomes, we make the same assumptions 
as in \citet{loehne-et-al-2007}.
This applies, in particular, to the critical energy for disruption and 
dispersal, $Q_D^*(s)$, as well as to the size distribution of fragments of
an individual collision. However, in contrast to \citet{loehne-et-al-2007}
where only catastrophic collisions were taken into account, we include here
cratering collisions as well. This is necessary, as cratering collisions
alter the size distribution of dust in the disk markedly, which shows up in
the SEDs \citep{thebault-et-al-2003,thebault-augereau-2007}.
The actual model of cratering collisions used here is close to that
by \citet{thebault-augereau-2007}.
An essential difference is our assumption of a single power law for
the size distribution of the fragments of an individual collision instead
of the broken power law proposed originally in \citet{thebault-et-al-2003}.
However, this difference has little effect on the resulting size distribution in
collisional equilibrium.

\subsection{Optical properties of dust}

An important issue is a choice of grain composition and morphology.
These affect both the dynamical model (through radiation pressure
efficiency
as well as bulk density)
and thermal emission model (through absorption efficiency).
Here we assume compact spherical grains composed of astronomical silicate
\citep[a.k.a. astrosilicate or astrosil,][]{laor-draine-1993}, similar to the MgFeSiO$_4$ olivine, 
with density of $3.3\g\cm^{-3}$.
Taking optical constants from \citet{laor-draine-1993}, we calculated
radiation pressure efficiency $Q_{pr}$ and absorption efficiency $Q_{abs}$ with a standard
Mie routine \citep{bohren-huffman-1983}.

\begin{figure}[h!]
  \begin{center}
  \includegraphics[scale=0.6]
  {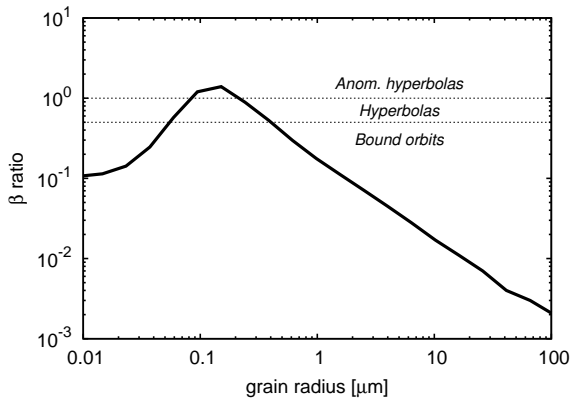}
  \caption
  {
  Radiation pressure to gravity ratio $\beta$ for astrosilicate grains
  as a function of their size.
  \corrected{Horizontal lines at $\beta = 0.5$ and $\beta = 1.0$ show,
  which particles typically move in bound elliptic orbits, in hyperbolas,
  as well as in anomalous hyperbolas (open outward from the star).}
  \label{fig_beta}
  }
  \end{center}
\end{figure}

To characterize the radiation pressure strength, it is customary to
use the radiation pressure to gravity ratio $\beta$ \citep{burns-et-al-1979},
which is independent of distance from the star and, for a given star,
only depends on $Q_{pr}$ and particle size. If grains that are small enough
to respond to radiation pressure derive from collisions of larger objects
in nearly circular orbits, they will get in orbits with eccentricities
$e \sim \beta / (1 - \beta)$. This implies that grains with $\beta < 0.5$ 
remain orbiting the star, whereas those with $\beta > 0.5$
leave the system in hyperbolic orbits.
The $\beta$ ratio for compact astrosil grains, computed from $Q_{pr}$,
is shown in Fig.~\ref{fig_beta}.
\corrected{The blowout limit, $\beta = 0.5$, corresponds to the grain
radius of $s = 0.4\mum$.}
Note that the tiniest astrosil grains ($ \la 0.1\mum$) would have $\beta < 0.5$ again
and thus could orbit the star in bound orbits.
However, the dynamics of these small motes would be
subject to a variety of effects (e.g. the Lorentz force) not included in our model,
and their lifetimes may be shortened by erosion processes (e.g. stellar wind sputtering).
Altogether, we expect them to make little contribution to the thermal emission in the mid-IR to
sub-mm. By setting the minimum radius of grains to $0.1\mum$, we
therefore do not take into account these grains here.

\begin{figure}[h!]
  \begin{center}
  \includegraphics[scale=0.6]
  {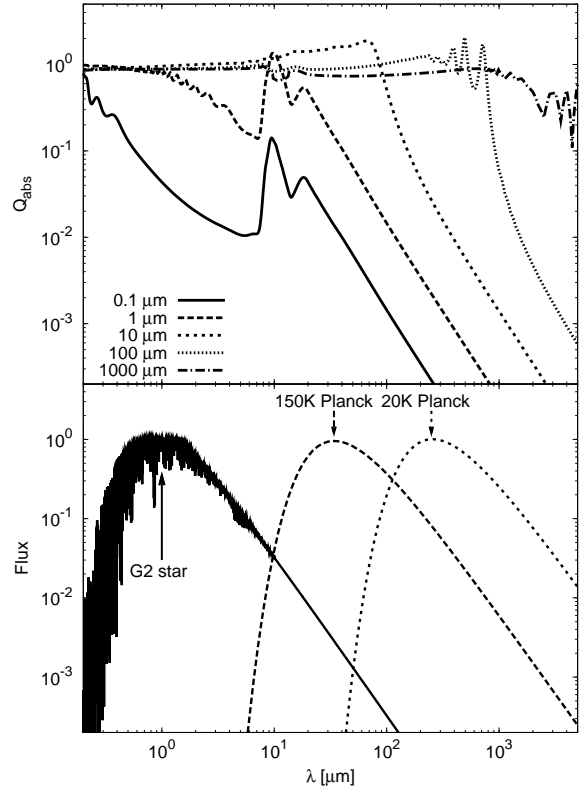}
  \caption
  {
Top: absorption efficiency of astrosilicate compact spherical grains as a function of
wavelength for different grain  sizes.
Bottom: the spectrum of a G2 V star and 
the Planck curves for 150 and 20 K (in arbitrary vertical scale)
to indicate the spectral ranges
most important for absorption and emission.
  \label{fig_Qabs}
  }
  \end{center}
\end{figure}

The spectral dependence of the absorption efficiency $Q_{abs}$ of different-sized
astrosil spheres is depicted in Fig.~\ref{fig_Qabs}.

\subsection{Parent planetesimal belts}

To have a representative set of ``reference'' debris disks 
around sun-like stars, we consider possible planetesimal rings centered
at the semimajor axes of $a = 3$, $10$, $30$, $100$, and $200\AU$ from the primary.
All five rings are assumed to have the same relative width initially
(again, in terms of semimajor axis) of $\Delta a /a = 0.2 $ ($\pm 0.1$)
and share the same semi-opening angle (the same as the maximum orbital inclination of the objects) of
$\varepsilon = 0.1$~rad. The orbital eccentricities of planetesimals
are then distributed uniformly between 0.0 and 0.2, in accordance with
the standard equipartition condition.
\corrected{The initial (differential) mass distribution of all solids is given
by a power law with the index $1.87$, a value that accounts for the
modification of the classical Dohnanyi's (\citeyear{dohnanyi-1969})
$1.833$ through the size dependence of material strength
\citep[see, e.g.,][]{durda-dermott-1997}.
}

\begin{deluxetable*}{rcccc}
\tablecaption{Description of reference disks
             \label{tab_runs}
             }
\tablewidth{0pt}
\tablehead{
           \colhead{Disk} &
           \colhead{Belt} &
           \colhead{Initial} &
           \colhead{$a$ range} &
           \colhead{$r$ range}\\
           \colhead{identifier} &
           \colhead{location [AU]} &
           \colhead{disk mass [$M_\oplus$]} &
           \colhead{[AU]} &
           \colhead{[AU]}
          }
\startdata
10EKBD \@@ \ph\ph  3AU &    3 &   0.001 &  0.3 --   30  &   0.5  --   20\\
10EKBD \@@ \ph    10AU &   10 &   0.03  &  1   --  100  &   2    --   50\\
10EKBD \@@ \ph    30AU &   30 &   1     &  3   --  300  &   5    --  200\\
10EKBD \@@       100AU &  100 &  30     & 10   -- 1000  &  20    --  500\\
10EKBD \@@       200AU &  200 & 200     & 20   -- 2000  &  30    -- 1000\\
\enddata
\end{deluxetable*}

The initial disk mass
is taken to be $1 M_\oplus$ (earth mass) for
a $30\AU$ ring, roughly corresponding to ten (or slightly more) times
the Edgeworth-Kuiper belt (EKB) mass \citep[e.g.][]{gladman-et-al-2001b,hahn-malhotra-2005}.
For other parent ring locations, the initial mass is taken in such a way
as to provide approximately the same spatial {\em density} of material.
Since the circumference of a ring $2 \pi a$,
its absolute width $\Delta a$,
and vertical thickness $2 a \varepsilon$
are all proportional to $a$,
the condition of a constant density requires the mass scaling $\propto a^3$.
This corresponds to the initial mass ranging from $\approx 0.001 M_\oplus$ in the $3\AU$ case
to $\approx 200 M_\oplus$ in the $200\AU$ case.
With these values, all reference disks have about ten times the
EKB density (10~EKBD).

\corrected{
That all the belts share the same volume density of material is
purely a matter of convention. Instead, we could choose them to have the same
surface density or the same total mass. Given the scaling rules, as discussed in the
text and Appendix~\ref{app_scaling}, none of these choices would have strong
advantages or disadvantages.
}

All five reference disks are listed in Table~\ref{tab_runs}.
We evolved them with the collisional code, ACE,
and stored all results between the ages of 10~Myr and 10~Gyr
at reasonable time steps.
In what follows, we use self-explanatory identifiers like 
\mbox{10EKBD \@@ 10AU \@@ 300Myr} to refer to
a particular disk of a particular age.

Importantly, the same runs of the collisional code automatically
provide the results for disks of any other initial density (or mass).
This is possible due to the mass-time scaling of \citet{loehne-et-al-2007},
which can be formulated as follows.
Denote by $M(M_0,t)$ the mass that a disk with initial mass $M_0$ has
at time $t$. Then, the mass of another disk with $x$ times larger initial
mass at time instant $t / x$ is simply
\be
       M(x M_0, t / x) = x M(M_0,t) ,
\label{mass_scaling}
\ee
For instance, the mass of the 1EKBD \@@ 10AU \@@ 10Gyr disk is one-tenth
of the 10EKBD \@@ 10AU \@@ 1Gyr disk mass.
Note that the same scaling applies to any other quantity directly proportional
to the amount of disk material.
In other words, $M$ may equally stand for the mass of dust,
its total cross section, thermal radiation flux, etc. 
See Appendix~\ref{app_scaling} for additional explanations.

\section{Results}

\subsection{Size and spatial distributions of dust}

As noted above, the collisional code ACE uses  masses and orbital elements
of disk particles as phase space variables. At any time instant, their
phase space distribution is transformed to usual mass/size and spatial distributions.
It is important to understand that mass/size distributions and spatial distributions
cannot, generally, be decoupled from each other. Grains of different sizes
have different radial distributions and conversely, the size distribution of
material is different at different distances from the star.

\begin{figure}[h!]
  \begin{center}
  \includegraphics[scale=0.6]
  {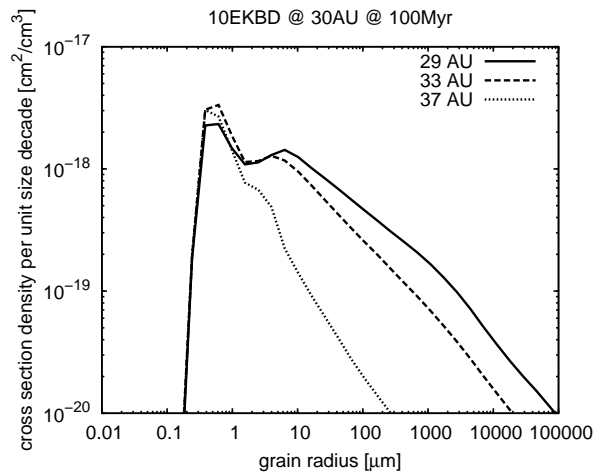}
  \caption
  {
  Size distribution in the 10EKBD \@@ 30AU \@@ 100Myr disk at three
  different distances from the star.
  \label{fig_size_dist}
  }
  \end{center}
\end{figure}

A typical size distribution of solids is shown in Fig.~\ref{fig_size_dist}
for one of the disks, namely for 10EKBD \@@ 30AU \@@ 100Myr.
Different lines correspond to different distances from the primary.
As expected, the size distribution is the broadest within the parent ring
of planetesimals. Farther out, it only contains grains which are small enough
to develop orbits with sufficiently large apocentric distances due to radiation 
pressure.

\begin{figure}[h!]
  \begin{center}
  \includegraphics[scale=0.6]
  {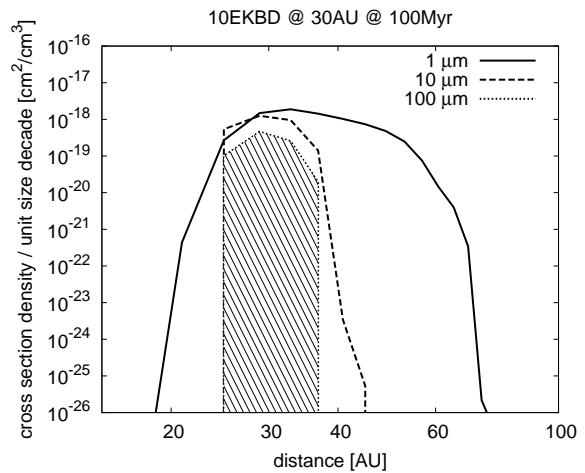}
  \caption
  {
  Spatial distribution of grains with three characteristic radii
  for the 10EKBD \@@ 30AU \@@ 100Myr disk. The ring of the biggest
  particles shown ($100\mum$, hatched) is the narrowest.
  Its radial extension is nearly the same as that of the initial
  planetesimal ring; vertical ``walls'' are artifacts due
  to a discrete distance binning.
  \label{fig_spat_dist}
  }
  \end{center}
\end{figure}

The spatial distribution of material in the same 
disk is shown in Fig.~\ref{fig_spat_dist}.
Here, different lines refer to different particle sizes.
The ring of the biggest particles shown ($100\mum$), for which radiation
pressure is negligible, nearly coincides with the initial ring
of planetesimals (semimajor axes: from $27$ to $33\AU$, eccentricities:
from 0.0 to 0.2, hence radial distances from $22$ to $40\AU$).
The larger the particles, the more confined their rings.
The rings are more extended outward with respect to the parent planetesimal ring
than inward.

\begin{figure}[h!]
  \begin{center}
  \includegraphics[scale=0.6]
  {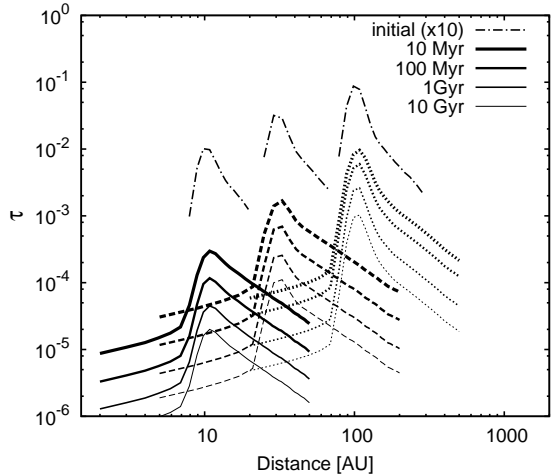}
  \caption
  {
  Radial profiles of the normal geometrical optical depth
  for three out of five basic runs
  (10EKBD \@@ 10AU, solid lines;
   10EKBD \@@ 30AU, dashed;
   10EKBD \@@ 100AU, dotted)
  at different ages.
  The thinner the line, the older the disk,
  as marked in the legend.
  The dashed-dotted lines are initial optical depths,
  artificially enhanced by a factor of ten for a better visibility.
  \label{fig_tau_dist}
  }
  \end{center}
\end{figure}

Radial profiles of the normal geometrical optical depth for
three reference disks (planetesimal rings at $10$, $30$, and $100\AU$) are depicted
in Fig.~\ref{fig_tau_dist}. Initially, the peak optical depth of the
disks is proportional to the distance of the parent ring, making
the $100\AU$ disk ten times optically thicker than the $10\AU$ one.
The subsequent collisional evolution of \corrected{the disks depends
on their initial mass and distance from the star, as explained
in detail in \citet{loehne-et-al-2007} and Appendix~\ref{app_scaling}.
Once a collisional steady state is reached (which is the case after 10~Myr
for all three disks),
the optical depth decays with time approximately as $t^\xi$,
where $\xi \approx {-0.3 \ldots -0.4}$,
i.e. roughly by one order of magnitude from 10~Myr to 10~Gyr.
In a steady-state regime, the optical depth is proportional
to $r^{1 + 1.3\xi} \sim r^{1.5}$. This explains why, at any age between
10~Myr and 10~Gyr, the $100\AU$ ring is $\approx 30$ times optically thicker
than the $10\AU$ one.
}

\subsection{Dust temperatures}

\begin{figure*}[hb!]
  \begin{center}
  \includegraphics[scale=0.6]
  {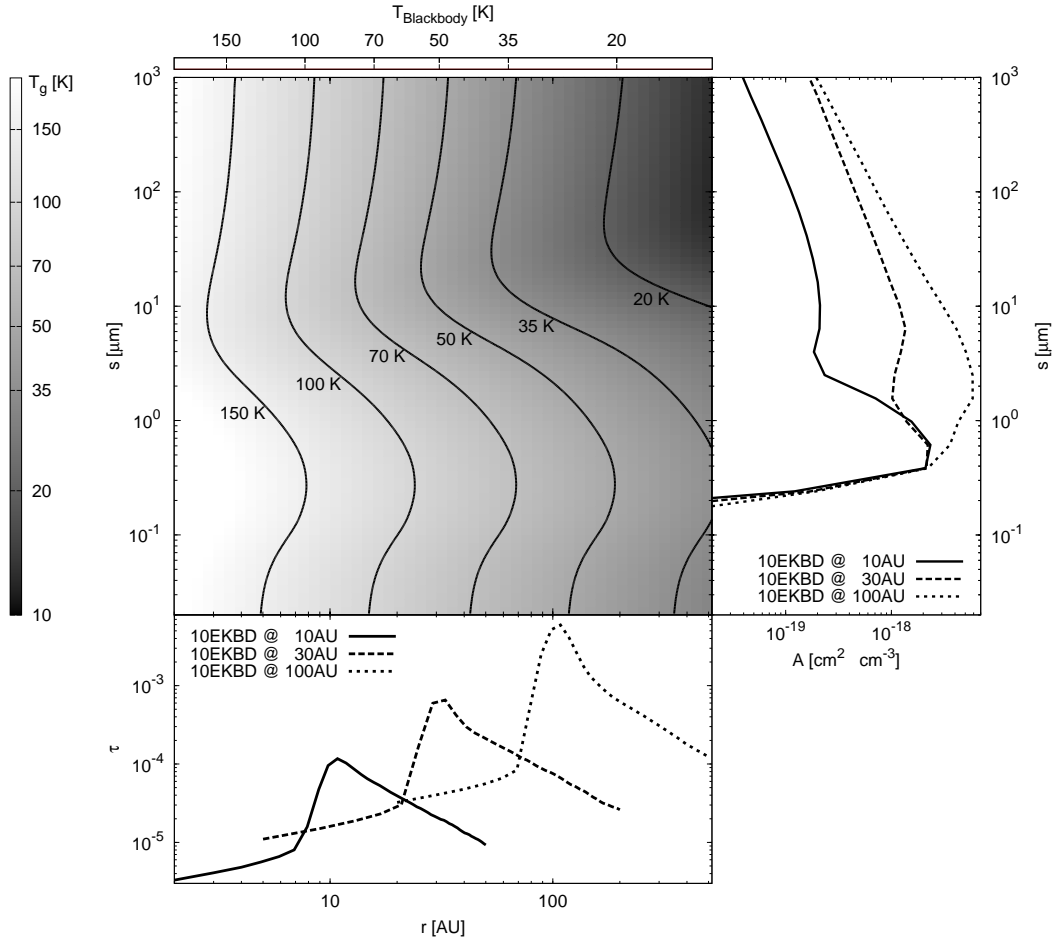}
  \caption
  {
The left upper plot shows the equilibrium temperature of dust particles as a function
of their distance from the star (horizontal axis) and size (vertical axis).
Contours are isotherms.
The blackbody dust temperatures are given along the upper edge of the plot for comparison.
In the right-hand plot the size distribution at the ``central'' distance of
the systems ($10\AU$, solid; $30\AU$, dashed;  and $100\AU$, dotted) 
at 100~Myr is given.
The lowest left plot gives the normal optical depth for the same three disks
as a function of distance to the star.
An intersection of a horizontal straight line going through the maximum
of the size distribution in a disk (right) with a vertical line through the peak
of its radial profile (bottom) provides the typical dust temperature in that disk.
  \label{fig_temp}
  }
  \end{center}
\end{figure*}

Figure~\ref{fig_temp} shows the dust temperatures as a function of two variables:
grain distances from the star and their radii.
In a parallel scale on the right,
we show typical size distributions (cf. Fig.~\ref{fig_size_dist}).
Similarly, under the temperature plot, typical radial profiles of the disk
are drawn (cf. Fig.~\ref{fig_spat_dist}). This enables a direct ``read-out''
of the typical\footnote{``Typical'' in the sense that it is the temperature
of cross-section dominating grains in the densest part of the disk.}
temperature in one or another disk. We find, for example,
$130\K$ at $10\AU$, $90\K$ at $30\AU$, and $50\K$ at $100\AU$.

\begin{figure}[h!]
  \begin{center}
  \includegraphics[scale=0.6]
  {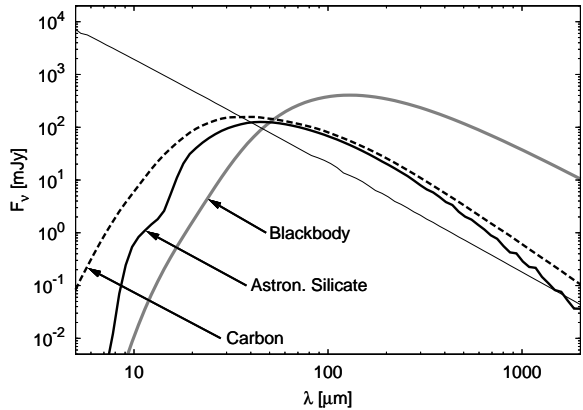}
  \caption
  {
  \rev{Figure changed.}
  The emission from one and the same, 1EKBD \@@ 30AU \@@ 100Myr, disk, calculated under
  different assumptions about absorbing and emitting properties of dust grains:
  blackbody, astrosil (our nominal case), and amorphous carbon particles.
  Thin solid line: photosphere of a G2V star.
  \label{fig_bb}
  }
  \end{center}
\end{figure}

\begin{figure*}
  \begin{center}
  \includegraphics[scale=0.6]
  {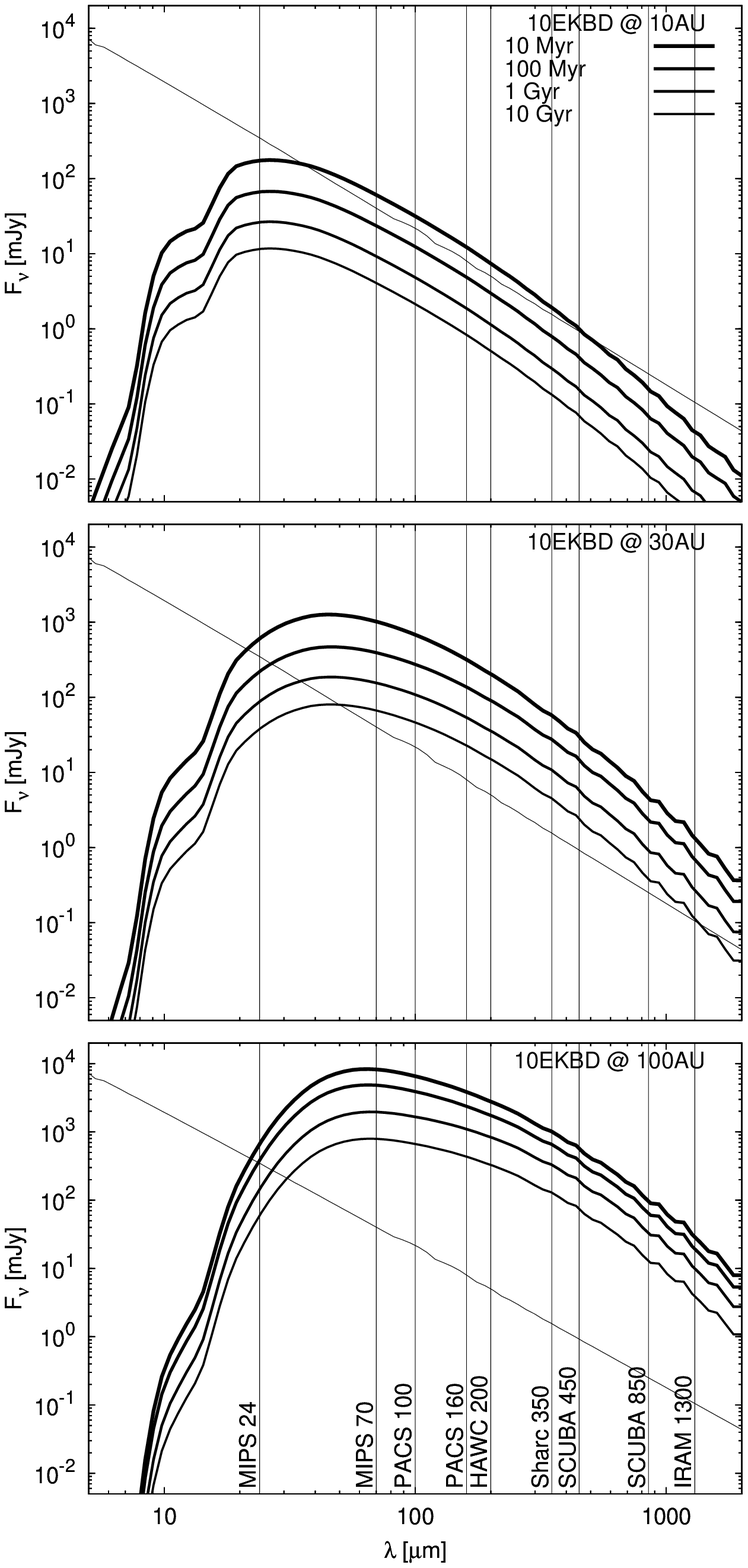}
  \includegraphics[scale=0.6]
  {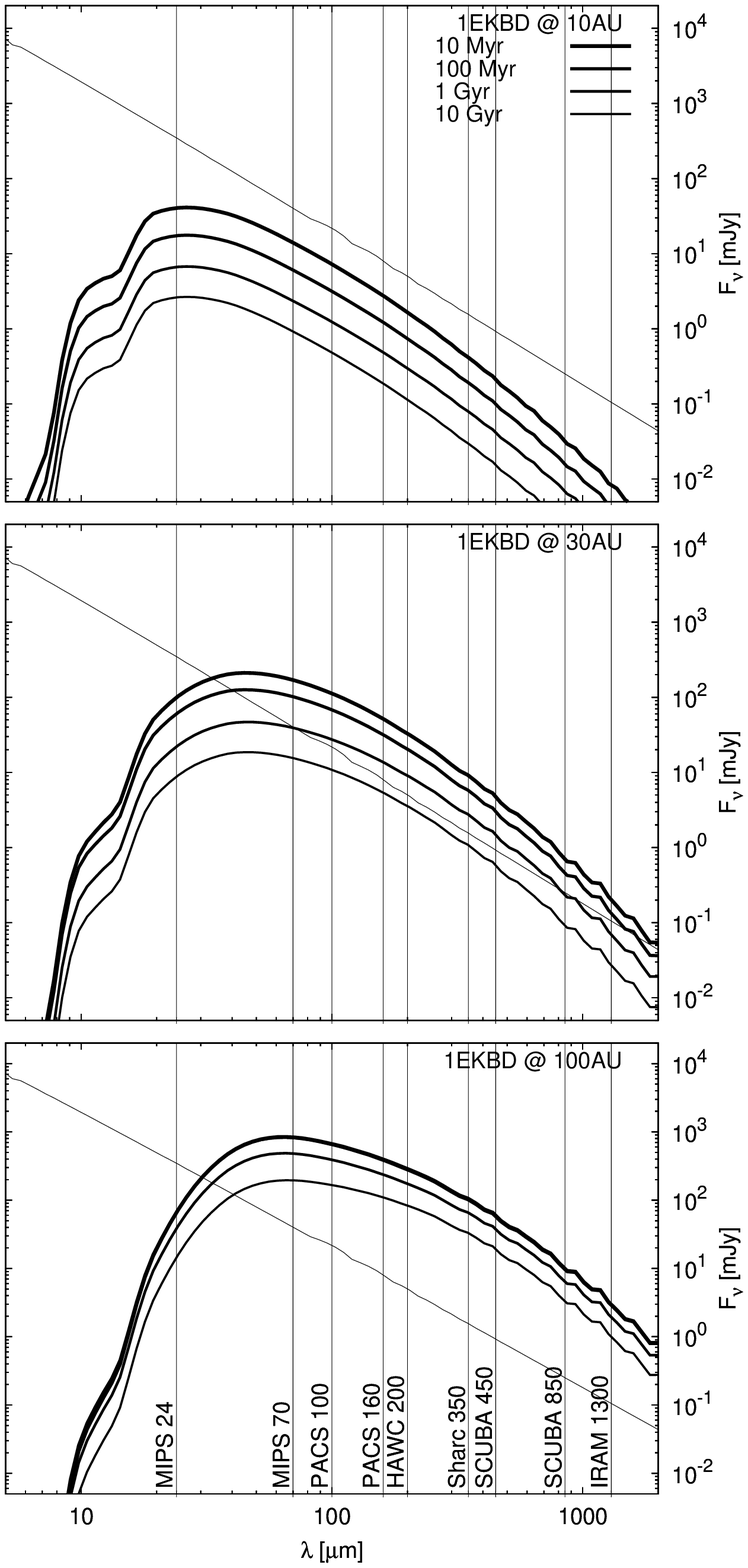}
  \caption
  {
  \rev{Some labels changed.}
Spectral energy distributions of disks
stemming from planetesimal rings with different masses at different locations
and  at different time steps.
To obtain the absolute values of fluxes, a distance of 10~pc was assumed.
Left: reference disks (10EKBD),
right: less massive disks (1EKBD).
The results for the latter have been obtained with the aid of 
Eq.~(\ref{mass_scaling}).
From top to bottom: the SEDs of the simulated planetesimal rings at 10, 30 and $100\AU$.
In each panel, lines of decreasing thickness correspond
to the ages of 10~Myr, 100~Myr, 1~Gyr, and  10~Gyr.
Note that the evolution of  the 1EKBD \@@ 100AU disk at the beginning is
very slow, so that the SEDs at 10 and 100~Myr are indistinguishable.
Vertical lines indicate centers of observational bands of several instruments (in $\mum$):
Spitzer MIPS        (24,  70, 160),
Herschel PACS       (100, 160),
\corrected{Sofia HAWC}          (200),
CSO Sharc           (350),
JCMT SCUBA/SCUBA 2  (450, 850),
\corrected{MPIfR IRAM}          (1300).
A thin line from top left to bottom right is the stellar photosphere.
  \label{fig_variation}
  }
  \end{center}
\end{figure*}

\begin{figure*}
  \begin{center}
  \includegraphics[scale=0.6]
  {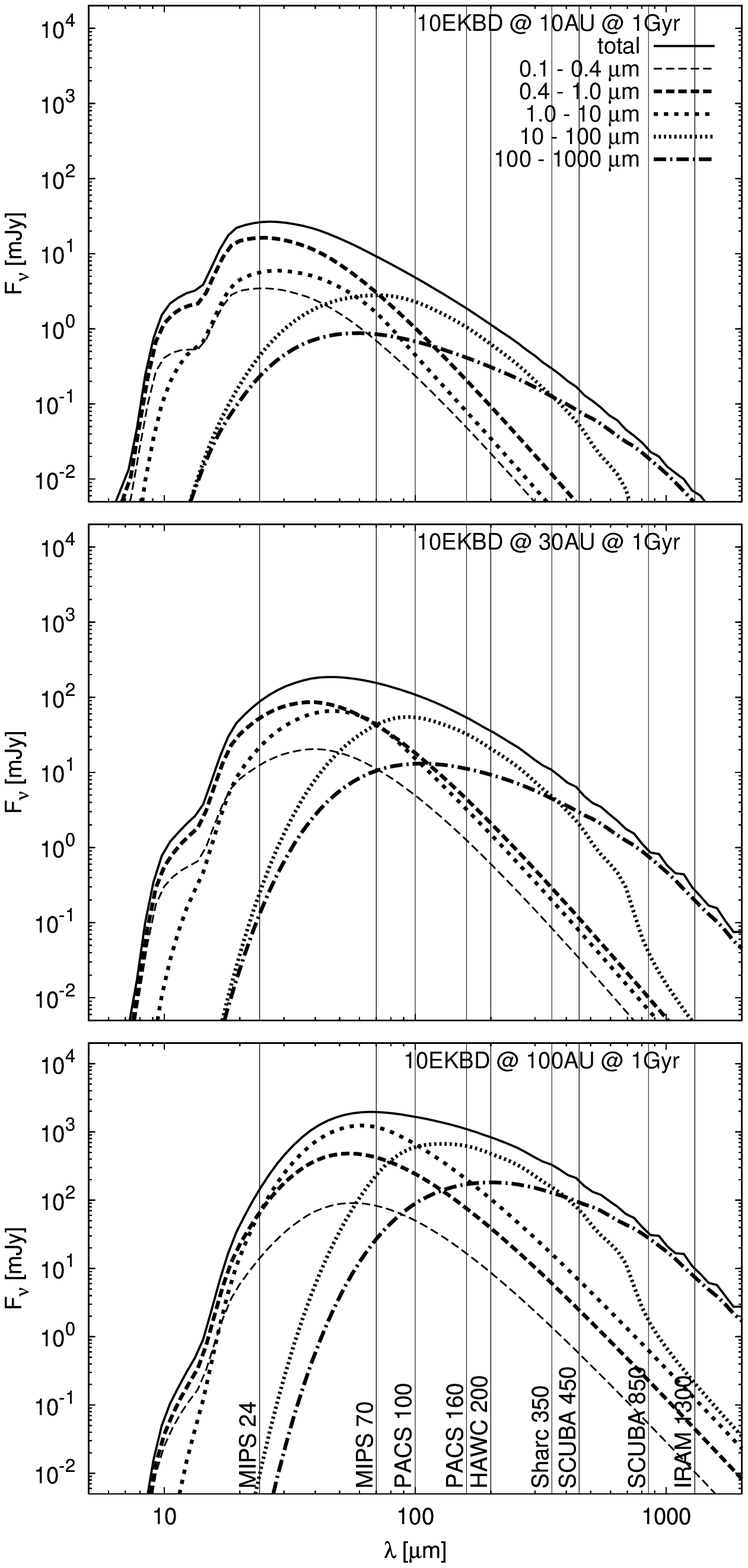}
  \includegraphics[scale=0.6]
  {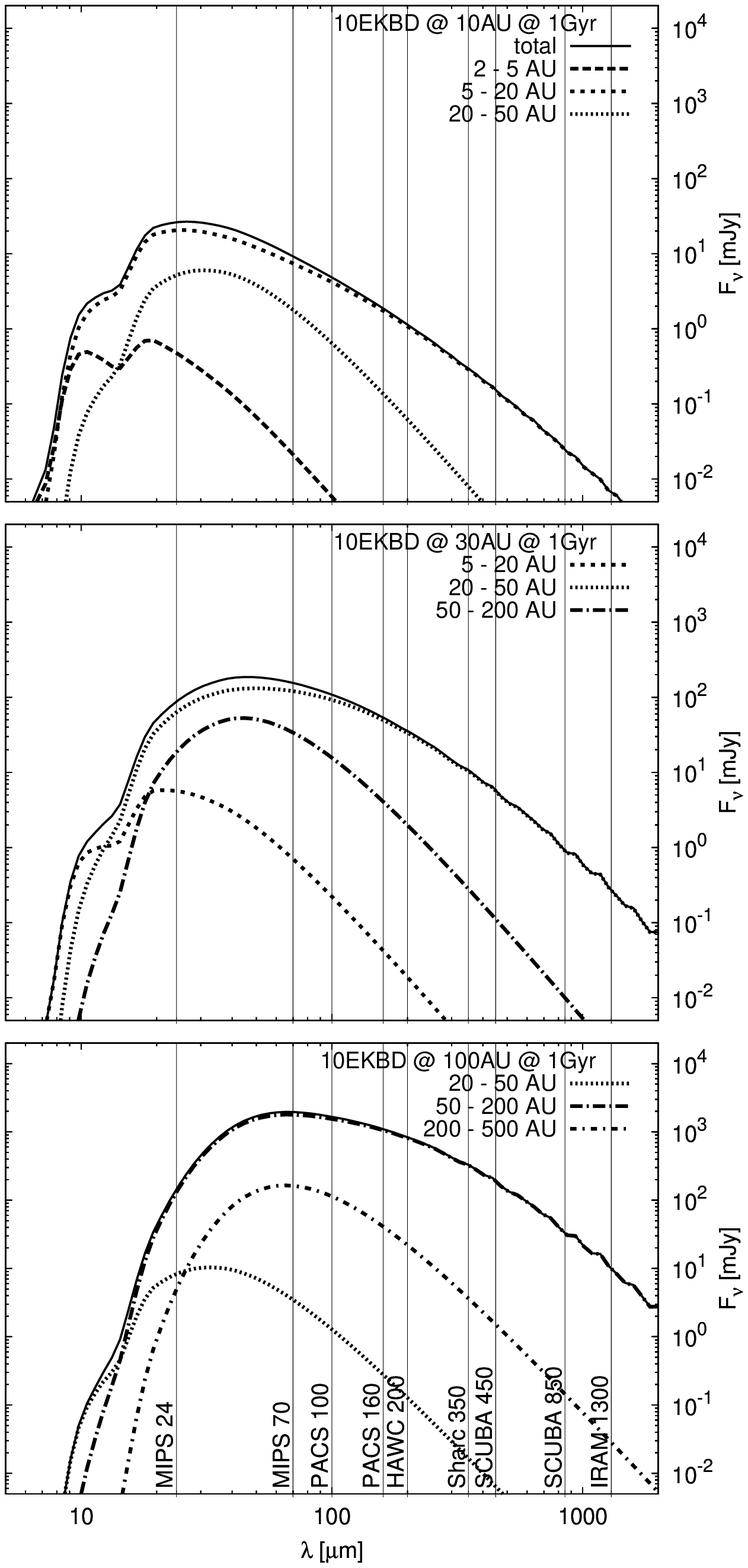}
  \caption
  {
  \rev{Figure changed.}
Contribution of individual grain size decades (shown with different linestyles in the
left panel) and individual radial annuli of the disks (different linestyles, right) to the SED.
\corrected{As the grain blowout radius is $\approx 0.4\mum$, see Fig.~\ref{fig_beta},
in the left panels we split the lowest
size decade into blowout grains with $s \in [0.1\mum, 0.4\mum]$ 
and bound ones with $s \in [0.4\mum, 1.0\mum]$.}
Panels from top to bottom correspond  to planetesimal rings at 10, 30 and $100\AU$.
The initial density of all disks is 10EKBD and their age is 1~Gyr.
  \label{fig_contribution}
  }
  \end{center}
\end{figure*}

These values are noticeably higher than the blackbody values
of $88\K$, $51\K$, and $28\K$, respectively.
The reason for these big deviations and for the S-shaped isotherms
in Fig.~\ref{fig_temp}
is the astronomical silicate's
spectroscopic properties with relatively high absorption at
visible wavelengths and steeply decreasing absorption coefficient
at \corrected{longer wavelengths} (see Fig.~\ref{fig_Qabs}).
The cross-section dominating astrosil grains are in a size range where the
absorption efficiency for visible and near-infrared wavelengths
(around $1\mum$) has already reached the blackbody value while
emission is still rather inefficient. With the enhancement of the
emission efficiencies relative to the ``saturated'' absorption,
temperatures drop drastically for somewhat larger grains.
The larger the distance from the star (yielding lower average temperature
and lower emission efficiency), the wider the size range
over which the temperature decreases, and the stronger the
temperature difference between small and large grains.
This explains why the S-shape of the isotherms gets more pronounced
from the left to the right in Fig.~\ref{fig_temp}.

Further, we note that Mie resonances
can increase the absorption/emission efficiencies even beyond unity
for wavelengths somewhat longer than the grain size (see $1\mum$, $10\mum$,
and $100\mum$ curves in Fig.~\ref{fig_Qabs}). This explains
the temperature maximum for grains of about $0.3\mum$ radius
(``resonance'' with the stellar radiation maximum) and the minimum
with temperatures even below the blackbody values for 10 to $50\mum$
grain radius (``resonance'' with the blackbody emission peak).

\subsection{Spectral energy distributions}

We start with a single, ``typical'' SED for one of the disks.
Such an SED for the 1EKBD \@@ 30AU \@@ 100Myr disk is shown in
Fig.~\ref{fig_bb} with a thick solid line.
It peaks at about $50\mum$, which is consistent with
the dust temperatures (Fig.~\ref{fig_temp}).
The hump at $\approx 10\mum$ is due to a classical silicate feature,
as discussed below.

For comparison, we have overplotted the SEDs calculated for the same disk,
but under different assumptions about the absorbing and emitting properties of grains:
in a black-body approximation \corrected{(grey line)}
and for amorphous carbon \corrected{(dashed line)}.
Note that the difference applies only to the calculation of thermal emission.
In other words, the dynamical modeling was still done by assuming the radiation pressure of astrosil
and not of perfectly absorbing  or carbon particles, but we assumed the grains to absorb
and emit like a blackbody or carbon when calculating the thermal emission. 
There is a striking difference between the curves, especially the blackbody SED
deviates from the others dramatically.
The blackbody assumption leads to a strong increase of the total flux
as well as to a shift of the
maximum in the SED from 50 to $130\mum$!
In addition the excess drops towards longer wavelengths much slower
than in the case of the astronomical silicate. In fact,
it will never intersect the stellar photospheric flux.

We now proceed with a set of SEDs for our grid of reference disks.
Some of them are shown in Fig.~\ref{fig_variation}.
The main features of these plots reveal no surprises.
The absolute level of excess emission is higher for more massive disks,
as well as for distant ones (which is just the consequence of the
assumed ``same-density'' scaling, as described in Sect. 3.5, see also
Fig.~\ref{fig_tau_dist}). The amount of dust emission is roughly comparable
with the photospheric emission for the mid-aged 1EKBD \@@ 30AU disk.
This is consistent with the known fact that a several Gyr-old EKB counterpart
would only slightly enhance the photospheric
emission even at the ``best'' wavelengths.
The position of the maximum emission ranges from $\approx 30\mum$ for the $10\AU$ disk
to $\approx 70\mum$ for the $100\AU$ disk. Note that blackbody calculation would
predict the emission to peak at longer wavelengths; beyond $100\mum$ for a $100\AU$ disk.

Again, the hump seen in all SEDs slightly below $10\mum$ is due to a silicate feature in $Q_{abs}$;
furthermore, some traces of the second feature at $20\mum$ are barely visible.
This explanation is supported by Fig.~\ref{fig_Qabs}
that shows the absorption efficiency feature in this spectral range for
small particles. This becomes even more obvious by comparing the contribution of the
different grain size decades. For $0.1$ to $1\mum$ particles the hump is more pronounced
than for larger ones (see left panels in Fig.~\ref{fig_contribution} below), as it is
the case for the absorption efficiency. Further on,
the $10\mum$ ``excess'' becomes less visible for most distant disks (from top to bottom
panels in Fig.~\ref{fig_variation}), where
the average temperatures are lower, the maximum emission shifts to longer wavelengths,
and therefore the Planck curve at $\lambda \sim 10\mum$--$20\mum$ is steeper.

The left panels in Fig.~\ref{fig_contribution} illustrate relative contributions of different-sized
particles to the full SEDs.
This is useful to get an idea which
instrument is sensitive to which grain sizes.
\corrected{The blowout grains with radii less than $0.4\mum$ make only modest contribution
to the flux even at $10\mum$.
The mid-IR fluxes are always dominated by bound grains with 
$0.4\mum$ to $1\mum$ radii (for the $10\AU$ and $30\AU$ rings)
or those with $0.4\mum$ to $10\mum$ (for the $100\AU$ ring).
In the far-IR, particles up to $100\mum$ in size play a role.
The greatest effect on the sub-mm fluxes is that of $100\mum$ to $1\mm$ grains.
}

The position of the different maxima \corrected{in Fig.~\ref{fig_contribution}}
can be understood by comparing the size decades to the dust temperature plot, Fig.~\ref{fig_temp}.
Particles of $0.1\mum$ to $1\mum$ are on the average a bit warmer than particles of $1$ to $10\mum$.
However, the size distribution shows that the second decade is dominated by particles only slightly
larger than $1\mum$, which are still nearly as warm as the particles in the decade below.
Thus, the maxima of the corresponding SED contributions are shifted only slightly.
It is the step to the next decade where the decrease of temperature becomes very obvious
by a large shift of the maximum. From that size on, the maxima stay nearly at the same
position (in fact the maxima are shifted again to smaller wavelengths) as the
temperature changes only marginally.

Similar to the contribution of the different size decades in the left panel,
the right panels in Fig.~\ref{fig_contribution} demonstrate
the contribution of the different radial parts of the disk to the total SED.
As expected, most of the flux comes from the medium distances
as this is the location of the birth ring. The second largest contribution
is made by the outer part of the ring.

\section{Application to selected debris disks}

\rev{This section was further split into several additional subsections}

\subsection{Measured fluxes}

\begin{figure*}[b!]
  \begin{center}
  \includegraphics[scale=0.6]
  {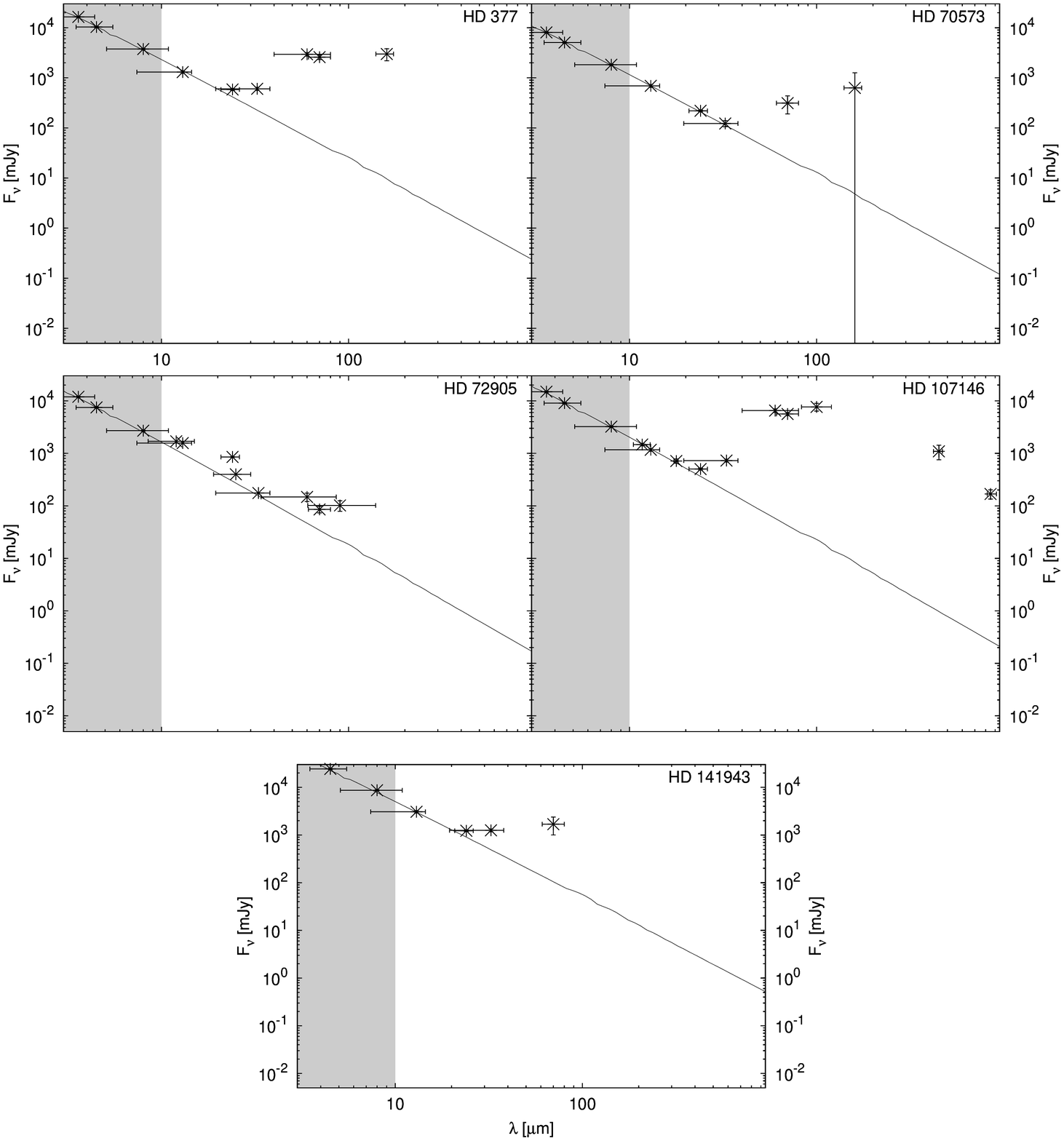}
  \caption
  {
  \rev{Figure changed.}
Observational data for five selected G2V stars.
Note that all fluxes have been scaled to the same standard distance of 10pc.
\corrected{Symbols} in the left-hand, \corrected{grey-shaded} part of each panel ($\lambda < 10\mum$)
are IRAC observations.
They are used to find an appropriate Hauschildt model to the photosphere
(thin solid line), assuming that no excess is already present in the near infrared.
\rev{Text moved to the caption of the next figure.}
Vertical error bars are $1\sigma$ observational uncertainties, taken from the source papers.
Horizontal bars indicate the band width of the
respective detector.
  \label{fig_obs}
  }
  \end{center}
\end{figure*}

To test the plausibility of our models, we have selected several nearby
sun-like stars known to possess debris dust. We used published datasets
to search for stars with (i) spectral classes most likely G2V (or very close),
and (ii) unambiguous excesses probed in a wide range of wavelengths
from near-IR to far-IR or sub-mm.
The resulting five stars and their properties are
listed in Table~\ref{tab_stars}, a summary of observational data on them
is given in Table~\ref{tab_obs},
and the disk properties as derived in original papers are collected in Table~\ref{tab_props_previous}.
The data include those from various surveys 
with IRAS, ISO, Spitzer, Keck II, and JCMT 
(Table~\ref{tab_obs}).
The estimated ages of the systems
range from 30 to 400~Myr
(Table~\ref{tab_stars})
and the fractional luminosities from $\sim 10^{-5}$ to $\sim 10^{-3}$
(Table~\ref{tab_props_previous}).
\corrected{The collected data points for our sample stars (photosphere $+$ dust)
are plotted in Fig.~\ref{fig_obs}.}

\begin{deluxetable}{l c c c c}
\tablecaption{Stellar parameters \label{tab_stars}}
\tablewidth{0pt}
\tablehead{
           \colhead{Star} &
           \colhead{$T_{\mathrm{eff}}$ [K]}&
           \colhead{$\log{L_*/L_\odot}$} &
           \colhead{D [pc]} &
           \colhead{age [Myr]}
          }
\startdata
      HD 377       & 5852 $^{a)}$            & 0.09 $^{a)}$         & 40 $^{a)}$    & 32 $^{a)}$ \\
      HD 70573     & 5841 $^{a)}$            & -0.23 $^{a)}$        & 46 $^{a)}$    & 100 $^{a)}$ \\
      HD 72905$^1$ & 5831 $^{a)}$            & -0.04 $^{a)}$        & 13.85 $^{d)}$ & 420 $^{d)}$ \\
      HD 107146    & 5859 $^{a)}$            & 0.04 $^{a)}$         & 29 $^{a)}$    & $100^{+100}_{-20}$ $^{c)}$ \\
      HD 141943    & 5805 $^{a)}$            & 0.43 $^{a)}$         & 67 $^{a)}$    & 32 $^{a)}$  \\
\enddata
\tablerefs{
$^{a)}$ \citet{hillenbrand-et-al-2008};
$^{b)}$ \citet{rhee-et-al-2007};
$^{c)}$ \citet{moor-et-al-2006}, \citet{trilling-et-al-2008}
          }
\tablecomments{$^1$A G1.5 star.}
\end{deluxetable}

\begin{deluxetable*}{l l l l}
\tablecaption{Observational data for the five G2 stars and their disks \label{tab_obs}}
\tablewidth{0pt}
\tablehead{
           \colhead{Star} &
           \colhead{Instrument, $\lambda$ ($\mum$)} &
           \colhead{Reference} &
           \colhead{Notes}
          }
\startdata
      HD 377     & {\it IRAC} 3.6/4.5/8.0   & \citet{hillenbrand-et-al-2008}   & \\
                 & {\it IRAS} 13/33         & \citet{hillenbrand-et-al-2008}   & \\
                 & {\it IRAS} 60            & \citet{moor-et-al-2006}          & \\
                 & {\it MIPS} 24/70/160     & \citet{hillenbrand-et-al-2008}   & \\ \hline
      HD 70573   & {\it IRAC} 3.6/4.5/8.0   & \citet{hillenbrand-et-al-2008}   & A planet host star \\
                 & {\it IRS} 13/33          & \citet{hillenbrand-et-al-2008}   & \citep{setiawan-et-al-2007} \\
                 & {\it MIPS} 24/70/160     & \citet{hillenbrand-et-al-2008}   & \\ \hline
      HD 72905   & {\it IRAC} 3.6/4.5/8.0   & \citet{hillenbrand-et-al-2008}   & \\
                 & {\it IRS} 13/33          & \citet{beichman-et-al-2006}      & \\
                 & {\it IRAS} 12/25         & \citet{spangler-et-al-2001}      & \\
                 & {\it ISOPHOT} 60/90      & \citet{spangler-et-al-2001}      & \\
                 & {\it MIPS} 24            & \citet{bryden-et-al-2006}        & \\
                 & {\it MIPS} 70            & \citet{hillenbrand-et-al-2008}   & \\ \hline
      HD 107146  & {\it IRAC} 3.6/4.5/8.0   & \citet{hillenbrand-et-al-2008}   & Resolved in V and I\\
                 & {\it LWS} 11.7/17.8  & \citet{williams-et-al-2004}          & bands \citep{ardila-et-al-2004},\\
                 & {\it IRS} 13/33          & \citet{hillenbrand-et-al-2008}   & at 350 and $450\mum$ \\
                 & {\it IRAS} 60/100        & \citet{moor-et-al-2006}          & \citep{williams-et-al-2004},\\
                 & {\it MIPS} 24/70         & \citet{hillenbrand-et-al-2008}   & and at $3\mm$\\
                 & {\it SCUBA} 450/850      & \citet{williams-et-al-2004}      &  \citep{carpenter-et-al-2005}\\ \hline
      HD 141943  & {\it IRAC} 3.6/4.5/8.0   & \citet{hillenbrand-et-al-2008}   & \\
                 & {\it IRS} 13/33          & \citet{hillenbrand-et-al-2008}   & \\
                 & {\it MIPS} 24/70         & \citet{hillenbrand-et-al-2008}   & \\ \hline
\enddata
\end{deluxetable*}

\begin{deluxetable*}{l c c c c}
\tablecaption{Previously derived disk properties \label{tab_props_previous}}
\tablewidth{0pt}
\tablehead{
           \colhead{Star} &
           \colhead{$T_{\mathrm{dust}}$ [K]}  &
           \colhead{$R_{\mathrm{dust}}$ [AU]} &
           \colhead{$M_{\mathrm{dust}}$ [$M_\oplus$]} & 
           \colhead{$L_{\mathrm{dust}}/L_*$}
          }
\startdata
      HD 377     & 58 $^{a),1}$            & 23 $^{a),5}$             & $3.98\times 10^{-4}$ $^{a),8}$   & $3.98\times 10^{-4}$ $^{a),11}$      \\
                 &                         &                          &                                  & $(4.0\pm0.3)\times 10^{-4}$ $^{f),12}$ \\ \hline
      HD 70573   & 41 $^{a),1}$            & 35 $^{a),5}$             & $2.0\times 10^{-5}$ $^{a),8}$    & $1.0\times 10^{-4}$ $^{a),11}$       \\ \hline
      HD 72905   & 103 $^{a),1}$           & 7 $^{a),5}$              & $1.58\times 10^{-6}$ $^{a),8}$   & $2.0\times 10^{-5}$ $^{a),11}$       \\
                 & $63 - 67$ $^{b),3}$     & $12.2 - 15.9$ $^{b),3}$  & $3.3\times 10^{-6}$ $^{b),3}$    & $2.9\times 10^{-5}$ $^{b),13}$       \\
                 & 123 $^{g),2}$           & 6.2 $^{g),5}$            &                                  & $(0.6-1.5)\times 10^{-5}$ $^{g),14}$ \\
                 &                         &                          &                                  & $1.6\times 10^{-5}$ $^{e),15}$       \\
                 &                         &                          &                                  & $2.8\times 10^{-4}$ $^{g),16}$       \\ \hline
      HD 107146  & 52 $^{a),1}$            & 30 $^{a),5}$             & $1.26\times 10^{-3}$ $^{a),8}$   & $4.94\times 10^{-4}$ $^{a),11}$      \\
                 &                         & $13.6 - >200$ $^{a),6}$  &                                  &                                      \\
                 &                         &                          & $3.2\times 10^{-7}$ $^{c),9}$    & ($9.2\pm0.9)\times 10^{-4}$ $^{f),12}$ \\
                 & 55 $^{d),2}$            & 29 $^{d),5}$             & $8.99\times 10^{-2}$ $^{d),10}$  & $9.5\times 10^{-4}$ $^{d),12}$       \\
                 & 51 $^{h),4}$            & $>31 - 150$ $^{h),7}$    & $0.1$ $^{h),4}$                  & $1.2\times 10^{-3}$ $^{h)}$          \\ \hline
      HD 141943  & 85 $^{a),1}$            & 18 $^{a),5}$             & $7.94\times 10^{-5}$ $^{a),8}$   & $1.58\times 10^{-4}$ $^{a),11}$      \\
                 &                         & $8.6 - 40$ $^{a),6}$     &                                  &                                      \\ \hline
\enddata
\tablerefs{
a) \citet{hillenbrand-et-al-2008},
b) \citet{beichman-et-al-2006},
c) \citet{carpenter-et-al-2005},
d) \citet{rhee-et-al-2007}, 
e) \citet{bryden-et-al-2006},
f) \citet{moor-et-al-2006},
g) \citet{spangler-et-al-2001},
h) \citet{williams-et-al-2004}
          }
\tablecomments{\scriptsize
$^{1}$ Color temperature ($33 - 70 \mu\mbox{m}$) from blackbody SED fitting.
$^{2}$ From SED fitting using a single temperature blackbody.
$^{3}$ From SED fitting using $10\mu\mathrm{m}$ silicate grains with a temperature
              profile following a power law (favored model in \citet{beichman-et-al-2006}).
$^{4}$ From single temperature SED fitting using a modified blackbody and a mass absorption
              coefficient $\kappa_{850} = 1.7~\mathrm{cm}^2/\mathrm{g}$.
$^{5}$ Derived from $T_{\mathrm{dust}}$ assuming blackbody (lower limit).
$^{6}$ Extended ring derived from blackbody SED fitting assuming a constant surface density.
$^{7}$ Inner border derived from SED fitting, outer border taken from resolved image.
$^{8}$ Derived from fractional luminosity for an average grain size of $<a> = 10\mu\mathrm{m}$
              and a density of $\rho = 2.5\mathrm{g}/\mathrm{cm}^3$.
$^{9}$ Derived for $T_{\mathrm{dust}} = 40~\mathrm{K}$ using a frequency dependent mass
              absorption coefficient.
$^{10}$ Derived from submillimeter observations using a dust opacity of $1.7~\mathrm{cm}^2/\mathrm{g}$ at $850~\mum$.
$^{11}$ Derived from $T_{\mathrm{dust}}$ and $R_{\mathrm{dust}}$ using Stefan-Boltzmann relation.
$^{12}$ $L_{\mathrm{dust}}/L_* = L_{\mathrm{IR}}/L_{\mathrm{bol}}$.
$^{13}$ $L_{\mathrm{dust}}$ obtained by integrating {\it IRS} spectrum ($10 - 34~\mu\mathrm{m}$) after
               extrapolation to $70~\mu\mathrm{m}$.
$^{14}$ $L_{\mathrm{dust}}$ is derived from the SED fitting and $L_*$ is obtained by integrating
               the corresponding Kurucz model.
$^{15}$ Minimum value, derived from the $70~\mu\mathrm{m}$ measurement.
$^{16}$ $L_*$ is the stellar bolometric luminosity and $L_{\mathrm{dust}}$ is the sum of the
               luminosities in each ({\it IRAS}) wavelength band with a correction (for longer wavelengths).
}
\end{deluxetable*}

\subsection{Observed excesses}

\begin{figure*}[b!]
  \begin{center}
  \includegraphics[scale=0.6]
  {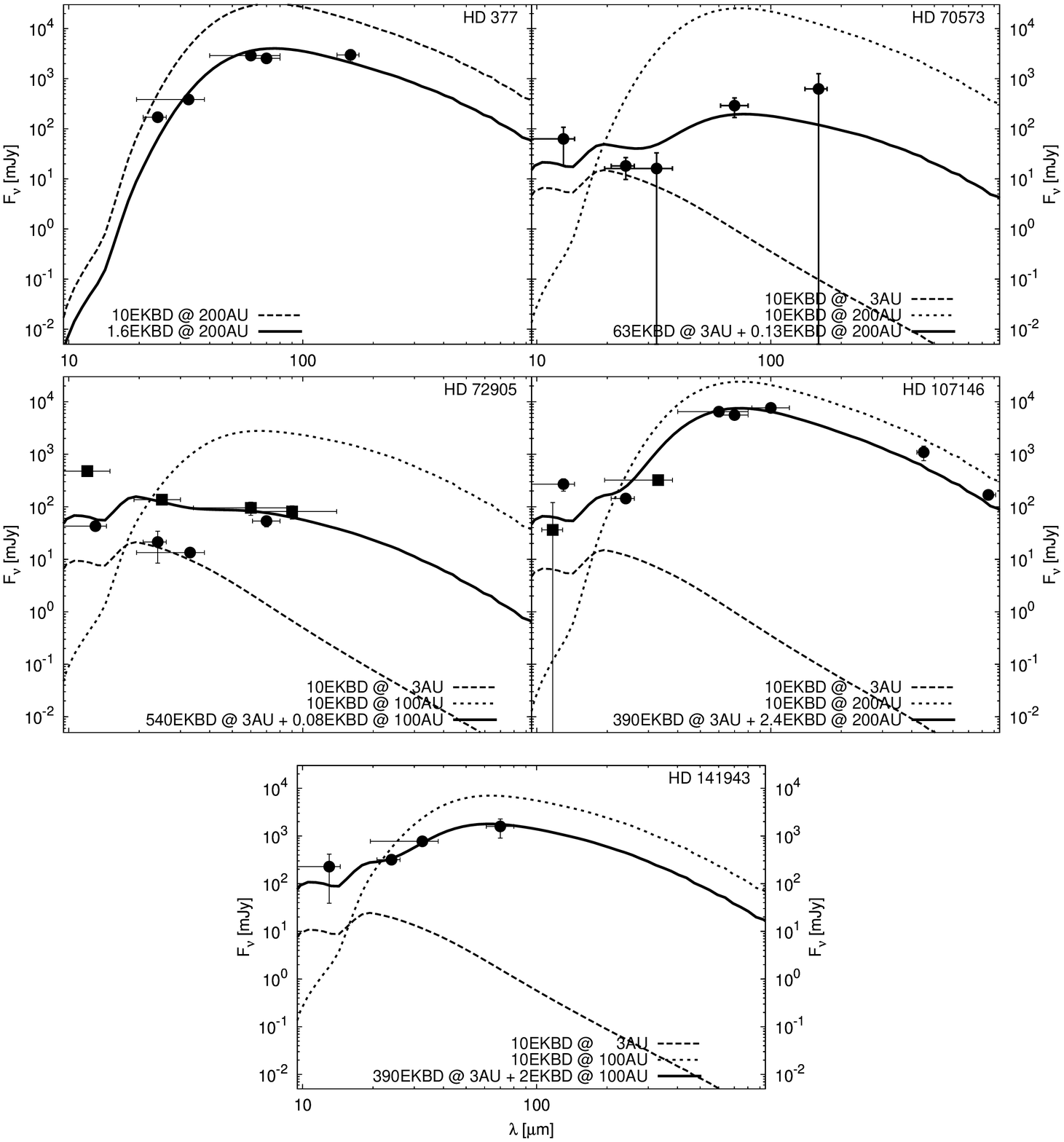}
  \caption
  {
\corrected{Observed (symbols) and modeled (lines) excess emission,
scaled to the distance of 10pc.
The wavelength range matches the unshaded part of Fig.~\ref{fig_obs}.
Here, in contrast to Fig.~\ref{fig_obs}, symbols represent the {\em excess} emission.
Squares mark the cases where the scaled NextGen model shown on that figure was used to
subtract the photosphere.
Circles indicate that for these observations the stellar emission was subtracted
using photospheric fluxes as given in the respective papers.
}
Dashed lines: two ``underlying'' SEDs of reference disks (unscaled, i.e. with 10EKBD),
one for ``cold'' excess and one for ``warm'' excess (except for HD~377 where only cold
component is observed).
Solid line: a linear combination of two scaled reference SEDs that provides
a reasonable fit to the data points (except for HD~377 where a single scaled reference
SED is sufficient).
  \label{fig_comp}
  }
  \end{center}
\end{figure*}

\corrected{Symbols in Fig.~\ref{fig_comp} represent}
the observed excess emission for our sample stars.
In the cases where the photospheric subtraction was done in the source papers, we just used the
published data points. In the cases where only the total measured flux (star + dust) was given,
we proceeded as follows.
Three IRAC points ($3.6$, $4.5$, and $8.0\mum$) were fitted by
an appropriate NextGen model \citep{hauschildt-et-al-1999}, and the resulting photospheric
spectrum was subtracted from the fluxes measured at longer wavelengths.
As far as the data quality is concerned,
the best case is clearly HD~107146, where
the data points cover a broad range between $10\mum$ and $1\mm$.
In other cases, the longest wavelengths probed lay at $70$--$160\mum$.
As a result, it \corrected{is} sometimes unclear where exactly the excess peaks.
This is exemplified by HD~70573 where the $160\mum$ point has a huge error bar. 

Yet before any comparison with the modeled SEDs,
the resulting points in \corrected{Fig.~\ref{fig_comp}} allow several quick conclusions.
Notwithstanding the paucity of long-wavelength data just discussed,
in all five systems the excess seems to peak at or slightly beyond $100\mum$,
suggesting a ``cold EKB'' as a source of dust. Additionally, in all systems 
except for HD~377, a warm emission at $\lambda <20\mum$ \corrected{seems to be} present,
implying a closer-in ``asteroid belt''.

\subsection{Comparison of measured and modeled SEDs}

We now proceed with a comparison between the observed dust emission
and the modeled emission.
We stress that our goal here is {\em not} to provide
the {\em best} fit to the observations possible with our approach, but rather to demonstrate
that a set of reference disks modeled in the previous sections can be used to make
rough preliminary conclusions about the planetesimal families.

To make such a comparison, we employ the following procedure:

1. For each star, we first look whether only cold or cold + warm excess emission
is present. In the former case (HD~377), we fit the data points with a single
``cold'' reference disk. In the latter case (all other systems), we invoke
a two-component model:
a close-in $3\AU$ disk and an appropriate ``cold'' disk.

2. The location of the ``cold'' planetesimal belt is chosen according
to the peak wavelength of the measured excess:
$100\AU$ (HD~72905 and HD~141943) or
$200\AU$ (HD~377, HD~70573, and HD~107146).

3. We then scale each of the two reference SEDs, ``warm'' and ``cold''
(or only one for HD~377) vertically to come to the observed absolute flux.
Physically, it necessitates a change in the initial disk mass.
However, it is {\em not} sufficient to
change the initial disk mass by the ratio of the observed flux and the flux from
a reference disk.
The reason is that a change in the initial mass also alters the rate
of the collisional evolution, whereas we need the ``right'' flux at a fixed
time instant, namely the actual age of the system (Tab.~\ref{tab_stars}).
Therefore, to find the mass modification factor we apply
scaling rules, as explained in Appendix~\ref{app_scaling}.
Specifically, we solve Eq.~({\ref{o-c}}).
In the systems that reveal both warm and cold emission, this is done separately
for the inner and outer disk.

The results presented in Fig.~\ref{fig_comp} \corrected{with lines}
show that the modeled SEDs can, generally, reproduce
the data points within their error bars.
Again, the judgment should take into account the fact that
we are just using one or two pre-generated SEDs for a rather coarse grid of
reference disks.
Much better fits would certainly be possible if we allowed a more exact positioning
of parent belts and let additional model parameters vary.
Dust opacities, initial distributions of planetesimals' \corrected{sizes and} orbital elements,
\corrected{as well as}
their mechanical properties that were fixed in modeling of the collisional outcomes
would all be at our disposal for this purpose. Further, more than two-component
planetesimal belts could be astrophysically relevant as well,
as is the case \corrected{in} our solar system
(asteroid belt, different cometary families, various populations in the EKB).

We now come to the interpretation of the fitting results, trying to recover
the properties of dust-producing planetesimal belts.
Table~\ref{tab_props} lists them for all systems.
The most important information is the deduced mass and location of the belts.

\begin{deluxetable*}{l l c c c c}
\tablecaption{Disk properties derived in this study \label{tab_props}}
\tablewidth{0pt}
\tablehead{
           \colhead{Star} &
           \colhead{Component} &
           \colhead{$M_{\mathrm{disk}}$ [$M_\oplus$] $^{1)}$} & 
           \colhead{$R_{\mathrm{belt}}$ [AU]         $^{2)}$} & 
           \colhead{$M_{\mathrm{dust}}$ [$M_\oplus$] $^{3)}$} & 
           \colhead{$T_{\mathrm{dust}}$ [K]          $^{4)}$} 
          }
\startdata
      HD \ph\ph\ph 377 & Outer &   $(32)$    \ph $32$      &  $200$     & $3.1\times 10^{-2}$ &  $40$      \\
      \hline
      HD \ph 70573     & Inner &   $(0.0063)$ \ph $0.0046$ &    $3$     & $1.4\times 10^{-7}$ & $200$      \\
                       & Outer &   $(2.6)$    \ph $2.5$    &  $200$     & $2.0\times 10^{-3}$ &  $40$      \\
      \hline
      HD \ph 72905     & Inner &   $(0.054)$ \ph $0.019$   &    $3$     & $3.4\times 10^{-8}$ & $200$      \\
                       & Outer &   $(0.23)$   \ph $0.23$   &  $100$     & $2.1\times 10^{-4}$ &  $50$      \\
      \hline
      HD 107146        & Inner &   $(0.039)$  \ph $0.023$  &    $3$     & $4.9\times 10^{-7}$ & $200$      \\
                       & Outer &   $(47)$    \ph $47$      &  $200$     & $4.8\times 10^{-2}$ &  $40$      \\
      \hline
      HD 141943        & Inner &   $(0.039)$  \ph $0.027$  &    $3$     & $8.0\times 10^{-7}$ & $200$      \\
                       & Outer &   $(6.1)$   \ph $6.1$     &  $100$     & $5.5\times 10^{-3}$ &  $50$      \\
\enddata
\tablecomments{
$^{1)}$ {\em Initial} mass (in parentheses) and the current mass of the whole planetesimal disk
(bodies up to $100\km$ in radius).\\
$^{2)}$ Location of the parent planetesimal belt.\\
$^{3)}$ Current mass of ``visible'' dust (grains up to $1\mm$ in radius).\\
$^{4)}$ Temperature of cross-section dominating astrosil grains at the location of the parent planetesimal belt,
see explanation at Fig.~\ref{fig_temp}.
}
\end{deluxetable*}

\subsection{Results for hot dust}

As far as the hot dust components in four out of five systems are concerned,
our results show that these can be explained by ``massive asteroid belts''
with roughly the lunar mass in bodies up to $\sim 100\km$ in size, located
at $3\AU$, with a width of $\sim 1\AU$.
However, the quoted distance of inner components~--- $3\AU$~--- is only due to
the fact that this is the smallest disk in our grid.
This distance can only be considered as an upper limit: the SEDs seem perfectly
compatible with disks as far in as $0.3\AU$, as suggested for the case of
HD~72905 \citep{wyatt-et-al-2007}.

What is more, even the very fact that hot excess is real can sometimes be questioned,
since it can be mimicked by photospheric emission slightly larger than the assumed values.
Indeed, the excess for HD~70573 and HD~72905 at wavelengths around and
below $25\mum$ does not exceed 10\%,
which is comparable with the average calibration uncertainty
and therefore has to be considered marginal \citep{bryden-et-al-2006,hillenbrand-et-al-2008}.
Only in the case of HD~72905, the Spitzer/IRS detection of the $10\mum$ emission from hot silicates
provides an independent confirmation that the hot excess is real \citep{beichman-et-al-2006}.
However, the HD~72905 plot in Fig.~\ref{fig_comp} makes it 
obvious that some problems occurred in terms of the photosphere fitting.
All data points that we obtained by subtracting the IRAC photospheric
fluxes (squares) systematically lie above the data points where a photosphere
from the literature was subtracted (circles).
The origin of the difference is unclear; on any account, the problem cannot
be mitigated by the assumption that an excess is already present at IRAC 
wavelengths, since this would shift the squares further upwards.
Considering the circles to be more trustworthy, the shape of the SED to
fit changes. Then a closer-in disk at $\sim 0.3\AU$ could better reproduce
the fluxes in the near and mid infrared, while
the outer ring would have to be shifted to a distance somewhat larger than
$100\AU$ in order not to surpass the measured flux at $33~\mum$.
A problem would arise with the inner disk: at $\sim 0.3\AU$, the collisional
evolution is so rapid that an unrealistically large initial belt mass would
be necessary.
Similar arguments have led \citet{wyatt-et-al-2007} to a conclusion that
HD~72905 must be a system at a transient phase rather than a system
collisionally evolving in a steady state.

Still, treating the derived sizes and masses of the inner disks as upper limits
yields physical implications.
Because the collisional evolution close to the star is rapid,
such belts must have lost up to two-thirds of their initial mass
before they have reached their present age
(cf. initial and current mass in Table~\ref{tab_props}).
In the case of HD~70573, the known giant planet with $a=1.76\AU$ and $e=0.4$
\citep{setiawan-et-al-2007}
does not seem to exclude the existence of a dynamically stable planetesimal
belt \corrected{either inside $\sim 1\AU$
or outside $\sim 3\AU$.}

\subsection{Results for cold dust}

The estimated parameters of the outer components of the disks suggest 
``massive and large Kuiper belts''.
The radii of the outer rings are larger
than the radii derived in previous studies
(cf. Table~\ref{tab_props_previous} and Table~\ref{tab_props}).
This traces back to our using astrosilicate instead of blackbody when calculating
the dust emission, so that the same dust temperatures are attained
at larger distances (see Fig.~\ref{fig_bb}).

Since one disk in our sample, that of HD 107146, has been resolved,
it is natural to compare our derived disk radius with the one obtained from the images.
\citet{williams-et-al-2004} report an outer border of the system of $150\AU$ based on submillimeter
images. In contrast, \citet{ardila-et-al-2004} detected an $85\AU$-wide ring peaking in
density at about $130\AU$. This is comparable to, although somewhat smaller than,
our $200\AU$ radius. However, moving the outer ring to smaller
distances would increase the fluxes in the mid infrared where the SED already surpasses the
observations and the other way round in the sub-mm region.
The resulting deficiency of sub-mm fluxes, though,
could be due to roughness of Mie calculations.
As pointed out by \citet{stognienko-et-al-1995},
an assumption of homogeneous particles typically leads to
underestimation of the amount of thermal radiation in the sub-mm region.

Large belt radii imply large masses.
Dust masses derived here are by two orders of magnitude larger
than previous estimates
(cf. Table~\ref{tab_props} and Table~\ref{tab_props_previous}).
The total masses of the belts we derive range
from several to several tens earth masses, to be compared
with $\sim 0.1 M_\oplus$ in the present-day EKB
(although there is no unanimity on that point~---
cf. \citeauthor{stern-colwell-1997} \citeyear{stern-colwell-1997}).
Note that, as the collisional evolution at
$100$--$200\AU$ is quite slow, whereas the oldest system in our sample is
only 420~Myr old, the difference between the initial disk mass and the current
disk mass is negligible.
Assuming several times the minimum mass solar nebula with a standard
surface density of solids $\Sigma \sim 50\g\cm^2 (r/1\AU)^{-3/2}$ \citep[e.g.][]{hayashi-et-al-1985},
the mass of solids in the EKB region would be a few tens $M_\oplus$;
and current models \citep[e.g.][]{kenyon-luu-1999b} successfully accumulate 100\km-sized
EKB objects in tens of Myr. However, it is questionable whether
the assumed radial surface density profile could extend much farther out
from the star.
As a result, it is difficult to say, whether a progenitor disk could contain enough
solids as far as at $200\AU$ from the star to form a belt of $30$ to $50 M_\oplus$.

However, such questions may be somewhat premature.
On the observational side, more data are needed, 
especially at longer wavelengths; for instance,
the anticipated Herschel data
(PACS at $100/160\mum$ and SPIRE at $250$ to $500\mum$)
would help a lot.
On the modeling side, a more systematic study is needed 
to clarify, how strongly various assumptions of the current model
\corrected{(especially the collisional outcome prescription and 
the material choices)} may affect the calculated size distributions
of dust, the dust grain temperatures, and the amount of their thermal emission.

At this point, we can only state that in the five systems
analyzed (with a possible exception of HD 72905)
and with the caveat that available data are quite scarce,
the observations are not incompatible with a standard steady-state
scenario of collisional evolution and dust production.
Of course, other possibilities, such as major collisional breakups
\corrected{\citep{kenyon-bromley-2005,grigorieva-et-al-2006}}
or events
similar to the Late Heavy Bombardment (as suggested, for instance for HD 72905,
\citeauthor{wyatt-et-al-2007} \citeyear{wyatt-et-al-2007})
cannot be ruled out for the inner disks.

\section{Summary}

Debris disks around main-sequence stars may serve as
tracers of planetesimal populations that have accumulated at earlier,
protoplanetary and transitional, phases of systems' evolution, and have not been
used up to form planets.
However, observations of debris disks are only sensitive to the lowest end of the size distribution.
Using dynamical and collisional models of debris disks is the only way 
to ``climb up'' the ladder of the collisional cascade, past the ubiquitous $\mu$m-sized
grains towards parent bodies and towards the main mass reservoir of the disks.

The main idea of this paper has been to take a grid of planetesimal families
(with different initial  masses, distances from a central star etc.),
to collisionally ``generate'' debris disks from these families
and evolve them with the aid of an elaborated  collisional code,
and finally, to calculate SEDs for these disks.
A comparison/fit of the observed SEDs with the pre-generated
SEDs is meant to allow quick conclusions about the properties of
the planetesimal belt(s) that maintain one or another observed disk.

Our specific results are as follows:

1. We have produced five reference disks around a G2V star from planetesimal belts at 3, 10, 30,
100, and $200\AU$ with 10 times the EKB mass density and evolved them for
10~Gyr. With an appropriate scaling rule (Eq.~\ref{scaling1}),
we can translate these results to an arbitrary initial disk mass
and any age between 10~Myr and 10~Gyr. Thus, effectively we have a
three-parametric set of reference disks (initial mass, location of
planetesimal belt, age). For all the disks, we have generated SEDs,
assuming astrosilicate (with tests made also for blackbody and amorphous carbon).

2. We have selected five G2V stars with good data (IRAS, ISO/ISOPHOT,
Spitzer/IRAC, /IRS, /MIPS, Keck II/LWS, and JCMT/SCUBA) and tested our grid against these
data. For all five systems, we have reproduced the data points within
the error bars with a linear combination of two disks from the grid
(an ``asteroid belt'' at $3\AU$ and an outer ``Kuiper belt'').
This automatically gives us the desired estimates of planetesimals
(location, total mass etc.).

3. A comparison of the observational data on the five stars with
the grid of models leads us to a conclusion that the cold emission
(with a maximum at the far-IR) is compatible with ``large Kuiper belts'',
with masses in the range 3--50 earth masses and radii of $100$--$200\AU$.
These large sizes trace back to the facts that the collisional model
predicts the observed emission to stem from micron-sized dust grains,
whose temperatures are well in excess of a blackbody temperature
at a given distance from the star
\citep[as discussed, e.g., in][]{hillenbrand-et-al-2008}.
This conclusion is rather robust against variation in parameters
of collisional and thermal emission models, and is roughly consistent
with disk radii revealed in scattered light images (e.g. HD~107146).
Still, quantitative conclusions about the
mass and location of the planetesimal belts would significantly
depend on (i) the adopted model of collision outcomes
\corrected{(which, in turn, depend on the dynamical excitation of
the belts, i.e. on orbital eccentricities and inclinations of planetesimals)}
and (ii) the assumed grains' absorption and emission efficiencies.
For example, a less efficient cratering
(retaining more grains with radii $\sim10\mum$ in the disk)
and/or more ``transparent'' materials 
(making dust grains of the same sizes at the same locations colder)
would result in ``shifting'' the parent belts closer to the star.

In future, we plan to extend this study in two directions.
First, we will investigate more systematically the influence of
the dust composition by trying relevant materials with available optical data
rather than astrosilicate; this should be done consistently in the
dynamical/collisional and thermal emission models.
Second, it is planned to extend this study to stars
with a range of spectral classes. This will result in a catalog of disk colors
that should be helpful for interpretation of data expected to come, most notably
from the Herschel Space Observatory.


\acknowledgements
This work has been particularly motivated by the Herschel Open Time Key Program
``DUNES'' (DUst disks around NEarby Stars, PI: C.Eiroa) and we wish to thank
many colleagues involved in DUNES (in particular, Jean-Charles Augereau,
Jens Rodmann, and Philippe Th\'ebault) for encouragement and numerous discussions.
\corrected{A speedy and constructive review of an anonymous reviewer helped to improve the paper.}
This research has been funded by the Deutsche Forschungsgemeinschaft (DFG),
projects Kr 2164/5-1 and Mu 1164/6-1, \corrected{by} the Deutscher Akademischer Austauschdienst 
(DAAD), project D/0707543,
\corrected{and by the International Space Science Institute (Bern)}.


\appendix

\rev{The Appendix has been reworked and extended.}

\section{Scaling rules}
\label{app_scaling}

{\em 1. Dependence of evolution on initial disk mass.}
Consider a disk with initial mass $M(t=0) \equiv M_0$ at a distance $r$ from the star
with age $t$.
Denote by $F(M_0,r,t)$ any quantity directly proportional
to the amount of disk material in any size regime, from dust grains to
planetesimals.
In other words, $F$ may equally stand for the total disk mass,
the mass of dust, its total cross section, etc. 
As found by \citet{loehne-et-al-2007}, there is a scaling rule:
\be
       F(x M_0, r, t) = x F(M_0, r, x t) ,
\label{scaling1}
\ee
valid for any factor $x > 0$. This scaling is an {\em exact} property of
every disk of particles, provided these are produced, modified and lost in
binary collisions and not in any other physical processes.

\bigskip
{\em 2. Dependence of evolution on distance.}
Another scaling rule is the dependence of the evolution timescale on
the distance from the star \citep{wyatt-et-al-2007,loehne-et-al-2007}.
Then
\be
       F(M_0, x r, t) \approx F(M_0, r, t^{-4.3}) .
\label{scaling2}
\ee
Unlike Eq.~(\ref{scaling1}), this scaling is approximate.

\begin{figure}[h!]
  \begin{center}
  \includegraphics[scale=0.6]
  {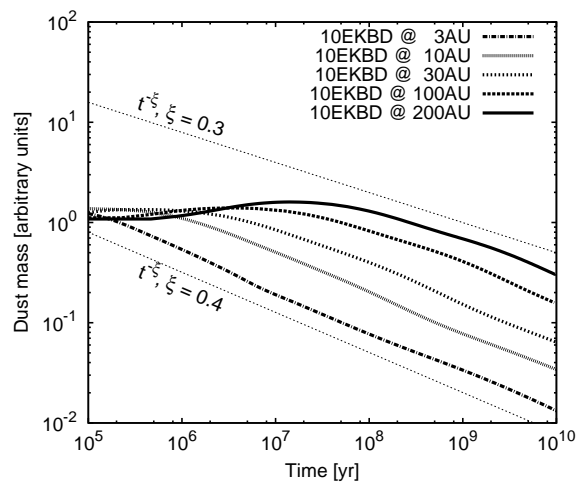}
  \caption
  {
  The time evolution of dust mass ($s< 1\mm$) for our five reference disks (thick lines).
  For comparison, power laws $t^{-\xi}$ with $\xi = 0.3$ and $\xi = 0.4$ are shown with thin dashed 
  lines.
  \label{fig_xi}
  }
  \end{center}
\end{figure}

\bigskip
{\em 3. Dust mass as a function of time.}
Finally, the third scaling rule found in 
\citet{loehne-et-al-2007}
is the power-law decay of the dust mass
\be
       F(M_0, r, x t) \approx x^{-\xi} F(M_0, r, t) ,
\label{scaling3}
\ee
where $\xi \approx 0.3 \ldots 0.4$ 
(Fig.~\ref{fig_xi}).
This scaling is also approximate and,
unlike Eq.~(\ref{scaling1}) and Eq.~(\ref{scaling2}), only applies
to every quantity directly proportional to the amount of {\em dust}.
In this context, ``dust'' refers to all objects in the strength rather than gravity
regime, implying radii less than about 100 meters.
The scaling is sufficiently accurate for disks that are much older than
the collisional lifetime of these $100\m$-sized bodies.
This is also seen in Fig.~\ref{fig_xi}: while for the $3\AU$ disk the power law (\ref{scaling3})
sets in after $\ll 1$~Myr, the $200\AU$ disk needs $\sim 100$~Myr to reach this regime.

Note that the ``pre-steady-state'' phase of collisional evolution may actually
require a more sophisticated treatment. Our runs assume initially a power-law size distribution
of planetesimals, and an instantaneous start of the collisional cascade at $t=0$.
In reality, an initial size distribution is set up by the accretion history of
planetesimals and will surely deviate from a single power law.
Moreover, at a certain phase cratering and destruction of objects may increasingly come into play
simultaneously with ceasing, yet ongoing accretion; the efficiencies and timescales of these 
processes will be different for different size ranges and different spatial locales in the disk
\citep[e.g.][]{davis-farinella-1997,kenyon-luu-1998,kenyon-luu-1999a,kenyon-luu-1999b}.

The usefulness of these scaling rules can be illustrated with the
following examples.

\bigskip
{\em Example~1.}
Assuming now $F$ to be the total amount of dust,
from Eqs.~(\ref{scaling1})--(\ref{scaling3}) one finds
\be
       F(xM_0, yr, t) \approx  x^{1-\xi} y^{4.3\xi}F(M_0, r, t) .
\ee
Our choice of reference disks (different distances, but the same volume density)
implies $x = y^3$.
The normal optical optical depth $\tau \propto F/r^2$ scales as
\be
       \tau(y^3 M_0, y r, t) \approx y^{1+1.3\xi} \tau (M_0, r, t) .
\ee
Therefore, once a steady-state is reached ($\xi \approx 0.3...0.4$),
a $y$ times more distant planetesimal belt gives rise to a
$y^{1+1.3\xi}$ times optically thicker disk.
This explains, in particular, why in Fig.~\ref{fig_spat_dist}
any $100\AU$ ring is $\approx 30$ times optically thicker
than the co-eval $10\AU$ one.

\bigskip
{\em Example~2.}
Since the distance $r$ in Eqs.~(\ref{scaling1}) and (\ref{scaling3})
is kept fixed, $F$ in these equations can also denote the radiation flux,
emitted by a disk at a certain wavelength.
Let $F_o(t)$ be the observed flux from a disk of age $t$.
Imagine a model of a disk of the same age with an initial mass $M_0$
predicts a flux $F(M_0,t)$ which is by a factor $A$ lower than the observed one:
\be
   F_o(t) = A F(M_0, r, t).
\ee
Our goal is to find the ``right'' initial mass, i.e. a factor $B$ such that
\be
   F_o(t) = F(B M_0, r, t).
\ee
With the aid of Eq.~(\ref{scaling1}), this can be rewritten as
\be
   F_o(t) = B F(M_0, r, B t) .
\label{o-c}
\ee
Eq.~(\ref{scaling3}) gives now
\be
   F_o(t)
   \approx B F(M_0, r, t) B^{-\xi} 
   = B ^{1-\xi} F(M_0, r, t) ,
\ee
whence
\be
  B \approx A^{1/(1-\xi)} .
\label{B(A)}
\ee
For instance, a 10 times higher flux at a certain age requires a 27--46 times larger
initial disk mass if $\xi =  0.3 \ldots 0.4$.

Although this rule is convenient for quick estimates, it should be used with caution.
As described above, the value of $\xi$ at the beginning of collisional evolution 
(which lasts up to 100~Myr for the $200\AU$ belt) can be much smaller~--- close to zero or even 
negative~--- than the ``normal'' $\xi =  0.3 \ldots 0.4$. For this reason, we prefer to use only
the first scaling rule, Eq.~(\ref{scaling1}). Therefore, instead of
applying Eq.~(\ref{B(A)}), we find $B$ by solving Eq.~(\ref{o-c}) numerically with a simple iterative 
routine.
It is this way Fig.~\ref{fig_comp} was constructed.



\begin{thebibliography}{46}
\expandafter\ifx\csname natexlab\endcsname\relax\def\natexlab#1{#1}\fi

\bibitem[{{Ardila} {et~al.}(2004){Ardila}, {Golimowski}, {Krist}, {Clampin},
  {Williams}, {Blakeslee}, {Ford}, {Hartig}, \&
  {Illingworth}}]{ardila-et-al-2004}
{Ardila}, D.~R.,
  et al.
  2004, \ApJ, 617, L147

\bibitem[{{Artymowicz}(1997)}]{artymowicz-1997}
{Artymowicz}, P. 1997, \ARevEPS, 25, 175

\bibitem[{{Aumann} {et~al.}(1984){Aumann}, {Beichman}, {Gillett}, {de Jong},
  {Houck}, {Low}, {Neugebauer}, {Walker}, \& {Wesselius}}]{aumann-et-al-1984}
{Aumann}, H.~H.,
  et al.
  1984, \ApJ, 278, L23

\bibitem[{Beichman {et~al.}(2005)Beichman, Bryden, Rieke, Stansberry, Trilling,
  Stapelfeldt, Werner, Engelbracht, Blaylock, Gordon, Chen, Su, \&
  Hines}]{beichman-et-al-2005}
  Beichman, C.~A.,
  et al.
  2005, \ApJ, 622, 1160

\bibitem[{{Beichman} {et~al.}(2006){Beichman}, {Tanner}, {Bryden},
  {Stapelfeldt}, {Werner}, {Rieke}, {Trilling}, {Lawler}, \&
  {Gautier}}]{beichman-et-al-2006}
  {Beichman}, C.~A.,
  et al.
  2006, \ApJ, 639, 1166

\bibitem[{Bohren \& Huffman(1983)}]{bohren-huffman-1983}
Bohren, C.~F., \& Huffman, D.~R. 1983, Absorption and Scattering of Light by
  Small Particles (Wiley and Sons: New York -- Chichester -- Brisbane --
  Toronto -- Singapore)

\bibitem[{{Bryden} {et~al.}(2006){Bryden}, {Beichman}, {Trilling}, {Rieke},
  {Holmes}, {Lawler}, {Stapelfeldt}, {Werner}, {Gautier}, {Blaylock}, {Gordon},
  {Stansberry}, \& {Su}}]{bryden-et-al-2006}
{Bryden}, G.,
  et al.
  2006, \ApJ, 636, 1098

\bibitem[{Burns {et~al.}(1979)Burns, Lamy, \& Soter}]{burns-et-al-1979}
Burns, J.~A., Lamy, P.~L., \& Soter, S. 1979, Icarus, 40, 1

\bibitem[{Carpenter {et~al.}(2005)Carpenter, Wolf, Schreyer, Launhardt, \&
  Henning}]{carpenter-et-al-2005}
Carpenter, J.~M., Wolf, S., Schreyer, K., Launhardt, R., \& Henning, T. 2005,
  \AJ, 129, 1049

\bibitem[{Davis \& Farinella(1997)}]{davis-farinella-1997}
Davis, D.~R., \& Farinella, P. 1997, Icarus, 125, 50

\corrected{
\bibitem[{Dohnanyi(1969)}]{dohnanyi-1969}
Dohnanyi, J.~S. 1969, \JGR, 74, 2531
}

\corrected{
\bibitem[{Durda \& Dermott(1997)}]{durda-dermott-1997}
Durda, D.~D., \& Dermott, S.~F. 1997, Icarus, 130, 140
}

\bibitem[{{Gladman} {et~al.}(2001){Gladman}, {Kavelaars}, {Petit},
  {Morbidelli}, {Holman}, \& {Loredo}}]{gladman-et-al-2001b}
{Gladman}, B.,
  et al.
  2001, \AJ, 122, 1051

\corrected{
\bibitem[{Grigorieva {et~al.}(2007)Grigorieva, Artymowicz, \&
  Th\'ebault}]{grigorieva-et-al-2006}
Grigorieva, A., Artymowicz, P., \& Th\'ebault, P. 2007, \AAp, 461, 537
}

\bibitem[{{Hahn} \& {Malhotra}(2005)}]{hahn-malhotra-2005}
{Hahn}, J.~M., \& {Malhotra}, R. 2005, \AJ, 130, 2392

\bibitem[{Hauschildt {et~al.}(1999)Hauschildt, Allard, \&
  Baron}]{hauschildt-et-al-1999}
Hauschildt, P., Allard, F., \& Baron, E. 1999, \ApJ, 512, 377

\bibitem[{{Hayashi} {et~al.}(1985){Hayashi}, {Nakazawa}, \&
  {Nakagawa}}]{hayashi-et-al-1985}
{Hayashi}, C., {Nakazawa}, K., \& {Nakagawa}, Y. 1985, in Protostars and
  Planets II, ed. D.~C. {Black} \& M.~S. {Matthews}, 1100--1153

\bibitem[{{Hillenbrand} {et~al.}(2008){Hillenbrand}, {Carpenter}, {Kim},
  {Meyer}, {Backman}, {Moro-Martin}, {Hollenbach}, {Hines}, {Pascucci}, \&
  {Bouwman}}]{hillenbrand-et-al-2008}
  {Hillenbrand}, L.~A.,
  et al.
  2008, \corrected{\ApJ, 677, 630}

\corrected{
\bibitem[{{Kenyon} \& {Bromley}(2005)}]{kenyon-bromley-2005}
{Kenyon}, S.~J., \& {Bromley}, B.~C. 2005, \AJ, 130, 269
}

\bibitem[{{Kenyon} \& {Luu}(1998)}]{kenyon-luu-1998}
{Kenyon}, S.~J., \& {Luu}, J.~X. 1998, \AJ, 115, 2136

\bibitem[{{Kenyon} \& {Luu}(1999{\natexlab{a}})}]{kenyon-luu-1999a}
---. 1999{\natexlab{a}}, \AJ, 118, 1101

\bibitem[{{Kenyon} \& {Luu}(1999{\natexlab{b}})}]{kenyon-luu-1999b}
---. 1999{\natexlab{b}}, \ApJ, 526, 465

\bibitem[{Krivov {et~al.}(2006)Krivov, L\"ohne, \&
  Srem\v{c}evi\'c}]{krivov-et-al-2006}
Krivov, A.~V., L\"ohne, T., \& Srem\v{c}evi\'c, M. 2006, \AAp, 455, 509

\bibitem[{Krivov {et~al.}(2000)Krivov, Mann, \& Krivova}]{krivov-et-al-2000b}
Krivov, A.~V., Mann, I., \& Krivova, N.~A. 2000, \AAp, 362, 1127

\bibitem[{Krivov {et~al.}(2005)Krivov, Srem\v{c}evi\'c, \&
  Spahn}]{krivov-et-al-2005}
Krivov, A.~V., Srem\v{c}evi\'c, M., \& Spahn, F. 2005, Icarus, 174, 105

\bibitem[{{Laor} \& {Draine}(1993)}]{laor-draine-1993}
{Laor}, A., \& {Draine}, B.~T. 1993, \ApJ, 402, 441

\bibitem[{L\"ohne {et~al.}(2008)L\"ohne, Krivov, \&
  Rodmann}]{loehne-et-al-2007}
L\"ohne, T., Krivov, A.~V., \& Rodmann, J. 2008, \ApJ, 673, 1123

\bibitem[{{Meyer} {et~al.}(2004){Meyer}, {Hillenbrand}, {Backman}, {Beckwith},
  {Bouwman}, {Brooke}, {Carpenter}, {Cohen}, {Gorti}, {Henning}, {Hines},
  {Hollenbach}, {Kim}, {Lunine}, {Malhotra}, {Mamajek}, {Metchev},
  {Moro-Martin}, {Morris}, {Najita}, {Padgett}, {Rodmann}, {Silverstone},
  {Soderblom}, {Stauffer}, {Stobie}, {Strom}, {Watson}, {Weidenschilling},
  {Wolf}, {Young}, {Engelbracht}, {Gordon}, {Misselt}, {Morrison}, {Muzerolle},
  \& {Su}}]{meyer-et-al-2004}
{Meyer}, M.~R., 
  et al.
  2004, \ApJS, 154, 422

\bibitem[{{Mo{\'o}r} {et~al.}(2006){Mo{\'o}r}, {{\'A}brah{\'a}m}, {Derekas},
  {Kiss}, {Kiss}, {Apai}, {Grady}, \& {Henning}}]{moor-et-al-2006}
{Mo{\'o}r}, A.,
  et al.
  2006, \ApJ, 644, 525

\bibitem[{{Najita} \& {Williams}(2005)}]{najita-williams-2005}
{Najita}, J., \& {Williams}, J.~P. 2005, \ApJ, 635, 625

\bibitem[{{Rhee} {et~al.}(2007){Rhee}, {Song}, {Zuckerman}, \&
  {McElwain}}]{rhee-et-al-2007}
{Rhee}, J.~H., {Song}, I., {Zuckerman}, B., \& {McElwain}, M. 2007, \ApJ, 660,
  1556

\bibitem[{Rieke {et~al.}(2005)Rieke, Su, Stansberry, Trilling, Bryden,
  Muzerolle, White, Gorlova, Young, Beichman, Stapelfeldt, \&
  Hines}]{rieke-et-al-2005}
Rieke, G.~H.,
  et al.
  2005, \ApJ, 620, 1010

\bibitem[{{Setiawan} {et~al.}(2007){Setiawan}, {Weise}, {Henning}, {Launhardt},
  {M{\"u}ller}, \& {Rodmann}}]{setiawan-et-al-2007}
{Setiawan}, J.,
  et al.
  2007, \ApJ, 660, L145

\bibitem[{{Siegler} {et~al.}(2007){Siegler}, {Muzerolle}, {Young}, {Rieke},
  {Mamajek}, {Trilling}, {Gorlova}, \& {Su}}]{siegler-et-al-2006}
{Siegler}, N.,
  et al.
  2007, \ApJ, 654, 580

\bibitem[{{Spangler} {et~al.}(2001){Spangler}, {Sargent}, {Silverstone},
  {Becklin}, \& {Zuckerman}}]{spangler-et-al-2001}
{Spangler}, C., {Sargent}, A.~I., {Silverstone}, M.~D., {Becklin}, E.~E., \&
  {Zuckerman}, B. 2001, \ApJ, 555, 932

\bibitem[{{Stern} \& {Colwell}(1997)}]{stern-colwell-1997}
{Stern}, S.~A., \& {Colwell}, J.~E. 1997, \ApJ, 490, 879

\bibitem[{{Stognienko} {et~al.}(1995){Stognienko}, {Henning}, \&
  {Ossenkopf}}]{stognienko-et-al-1995}
{Stognienko}, R., {Henning}, T., \& {Ossenkopf}, V. 1995, \AAp, 296, 797

\bibitem[{{Su} {et~al.}(2006){Su}, {Rieke}, {Stansberry}, {Bryden},
  {Stapelfeldt}, {Trilling}, {Muzerolle}, {Beichman}, {Moro-Martin}, {Hines},
  \& {Werner}}]{su-et-al-2006}
{Su}, K.~Y.~L.,
  et al.
  2006, \ApJ, 653, 675

\bibitem[{{Th{\' e}bault} \& {Augereau}(2007)}]{thebault-augereau-2007}
{Th{\' e}bault}, P., \& {Augereau}, J.-C. 2007, \AAp, 472, 169

\bibitem[{Th\'ebault {et~al.}(2003)Th\'ebault, Augereau, \&
  Beust}]{thebault-et-al-2003}
Th\'ebault, P., Augereau, J.-C., \& Beust, H. 2003, \AAp, 408, 775

\bibitem[{{Trilling} {et~al.}(2008){Trilling}, {Bryden}, {Beichman}, {Rieke},
  {Su}, {Stansberry}, {Blaylock}, {Stapelfeldt}, {Beeman}, \&
  {Haller}}]{trilling-et-al-2008}
{Trilling}, D.~E.,
  et al.
  2008, \ApJ, 674, 1086

\bibitem[{{Trilling} {et~al.}(2007){Trilling}, {Stansberry}, {Stapelfeldt},
  {Rieke}, {Su}, {Gray}, {Corbally}, {Bryden}, {Chen}, {Boden}, \&
  {Beichman}}]{trilling-et-al-2007}
{Trilling}, D.~E.,
  et al.
  2007, \ApJ, 658, 1289

\bibitem[{Williams {et~al.}(2004)Williams, Najita, Liu, Bottinelli, Carpenter,
  Hillenbrand, Meyer, \& Soderblom}]{williams-et-al-2004}
Williams, J.~P.,
  et al.
  2004, \ApJ, 604, 414

\bibitem[{{Wolf} \& {Hillenbrand}(2003)}]{wolf-hillenbrand-2003}
{Wolf}, S., \& {Hillenbrand}, L.~A. 2003, \ApJ, 596, 603

\bibitem[{Wyatt(2005)}]{wyatt-2005}
Wyatt, M.~C. 2005, \AAp, 433, 1007

\bibitem[{{Wyatt} {et~al.}(2007){Wyatt}, {Smith}, {Greaves}, {Beichman},
  {Bryden}, \& {Lisse}}]{wyatt-et-al-2007}
{Wyatt}, M.~C.,
  et al.
  2007, \ApJ, 658, 569

\end{thebibliography}

\end{document}